\documentclass[12pt,english]{article}
\usepackage[T1]{fontenc}
\usepackage[latin1]{inputenc}
\usepackage{a4wide}
\usepackage{graphicx}
\usepackage{amssymb}

\makeatletter

\providecommand{\tabularnewline}{\\}

\usepackage{graphicx}
\makeatletter
\renewcommand{\theequation}{\hbox{\normalsize\arabic{section}.\arabic{equation}}}
\@addtoreset{equation}{section}
\renewcommand{\thefigure}{\hbox{\normalsize\arabic{section}.\arabic{figure}}}
\@addtoreset{figure}{section}
\renewcommand{\thetable}{\hbox{\normalsize\arabic{section}.\arabic{table}}}
\@addtoreset{table}{section}
\makeatother 

\usepackage{babel}
\makeatother
\begin{document}

\title{{\normalsize\begin{flushright}ITP Budapest Report No. 613\end{flushright}}\vspace{1cm}(Semi)classical
analysis of sine-Gordon theory on a strip}

\author{Z. Bajnok$^{1}$%
\footnote{bajnok@afavant.elte.hu%
}, L. Palla$^{2}$%
\footnote{palla@ludens.elte.hu%
}, and G. Takács$^{1}$%
\footnote{takacs@ludens.elte.hu%
}\\
 $^{1}$\emph{\small Theoretical Physics Research Group, Hungarian
Academy of Sciences, }\\
\emph{\small 1117 Budapest, Pázmány Péter sétány 1/A, Hungary}\\
$^{2}$\emph{\small Institute for Theoretical Physics, Eötvös University
}\\
\emph{\small 1117 Budapest, Pázmány Péter sétány 1/A, Hungary}}

\maketitle
\begin{abstract}
Classical sine-Gordon theory on a strip with integrable boundary conditions
is considered analyzing the static (ground state) solutions, their
existence, energy and stability under small perturbations. The classical
analogue of Bethe-Yang quantization conditions for the (linearized)
first breather is derived, and the dynamics of the ground states is
investigated as a function of the volume. The results are shown to
be consistent with the expectations from the quantum theory, as treated
in the perturbed conformal field theory framework using the truncated
conformal space method and thermodynamic Bethe Ansatz. The asymptotic
form of the finite volume corrections to the ground state energies
is also derived, which must be regarded as the classical limit of
some (as yet unknown) Lüscher type formula. 
\end{abstract}

\section{Introduction}

Sine-Gordon theory is known as the prime example of integrable field
theory, and has been the subject of many investigations. In particular,
it is well understood on the half line with an integrable boundary
potential (or boundary condition of the Ghoshal-Zamolodchikov type
\cite{GZ}).

In the case of two-dimensional (mostly integrable) theories it is
of much interest to consider their dynamics in finite spatial volume
(strip or circle), since finite size effects (following the ideas
put forward by Lüscher \cite{Lusch} for the case of generic quantum
field theories) provide a very interesting link between different
descriptions of the system. Integrability, being based on the existence
of local conserved quantities, is generally preserved by passing from
the infinite volume system to a circle, or to a strip with boundary
conditions at the two ends that are both integrable in a semi-infinite
system. Besides that, sine-Gordon theory on the strip is also interesting
as a description of certain condensed matter systems \cite{R,LR,CSS1,CSS2,JS}.

In the case of sine-Gordon theory with boundaries, such investigations
so far mainly aimed to connect the exact scattering matrices derived
from the bootstrap (infrared data) to the perturbed conformal field
theory description (ultraviolet data). The main tools used to establish
this relation are Bethe Ansatz based (TBA, NLIE) or truncated conformal
space (TCS). While there is a great wealth of information provided
by these approaches, it is of interest to examine the dynamics of
the system from an intuitive, field-theoretic point of view, which
takes the classical Lagrangian as its starting point and is the goal
of the present work.

This approach starts with the classification of classical ground states
(vacua) of the system, which are identified with static solutions
of the field equations with appropriate boundary conditions at the
ends of the strip. This is followed by an investigation of the stability
of these states under linearized oscillations of the field. It is
very important to realize that beyond providing information on stability,
these small oscillations can be associated with the classical limit
of the elementary particle of sine-Gordon theory, which is the first
breather, therefore their examination gives information on the classical
limit of one-particle states containing a first breather on the strip.
In addition, the stability analysis forms the basis of semiclassical
quantization of the ground states on the strip (similarly to the case
of periodic boundary conditions discussed in \cite{MRS}), but we
do not attempt to solve this problem in this paper. Instead, we shall
be mainly interested in the relation between the very intuitive, visual
picture of the finite volume dynamics provided by this approach to
the more abstract, algebraic description provided by Bethe Ansatz
and TCS methods.

The outline of the paper is as follows. In Section 2 we overview some
general facts about the semi-infinite system, in particular the semiclassical
quantization of ground states discussed in \cite{KP}, which are needed
later. Section 3 outlines general properties of the system on a strip,
especially concerning the large volume limit. In Section 4 we present
a systematic investigation of the case with Dirichlet boundary conditions
at both ends of the strip, since it allows us to clarify all conceptual
issues in a very transparent way. In Section 5 we present another
case with a Dirichlet boundary condition at one end and a special
perturbed Neumann boundary condition at the other. Armed with all
the lessons we learned from these two cases, Section 6 presents a
discussion of the generic case. Conclusions are presented in Section
7, while two technical issues are relegated to appendices: Appendix
A contains the description and parameterization of the classical limit
of the reflection matrix of the first breather, and Appendix B contains
some numerical work on the infrared limit of the energy levels.

\section{Sine-Gordon theory on the half line }

The sine-Gordon theory in the bulk is described by the Lagrange density:
\[
\mathcal{L}_{SG}(x,t)=\frac{1}{2}\left(\partial_{\mu}\Phi(x,t)\right)^{2}-V\left(\Phi(x,t)\right)\]
 where the potential is: \[
V\left(\Phi\right)=\frac{m^{2}}{\beta^{2}}\left(1-\cos\beta\Phi\right)=2\frac{m^{2}}{\beta^{2}}\sin^{2}\frac{\beta\Phi}{2}\geq0\]
 Consider the restriction of this theory onto the negative half line,
with the (bulk) action\[
\mathcal{A}=\int_{-\infty}^{\infty}dt\int_{-\infty}^{0}dx\;\mathcal{L}_{SG}(x,t)\quad.\]
 Integrable boundary conditions correspond either to fixing the value
of the field at the origin (Dirichlet BC) \[
\Phi(x,t)|_{x=0}=\Phi_{0}^{D}\]
 or introducing a boundary potential of the form (perturbed Neumann
BC) \begin{equation}
V_{B}^{(z)}\left(\Phi(z,t)\right)=M_{z}\left[1-\cos\frac{\beta}{2}\left(\Phi(z,t)-\Phi_{z}\right)\right]\qquad,\qquad z=0\quad.\label{VB}\end{equation}
 The Dirichlet BC with $\Phi_{0}^{D}=\Phi_{0}$ can be recovered from
it by taking the $M_{0}\rightarrow\infty$ limit.

The semiclassical analysis of the static solutions in the perturbed
Neumann case was performed in \cite{KP}. Although all the results
for the Dirichlet case can be obtained from it in the $M_{0}\rightarrow\infty$
limit for pedagogical reasons, and because the explicit expressions
are needed later, we spell out the details here and refer for the
general case to \cite{KP}.

Using the various bulk symmetries of the model \cite{genpap} the
fundamental range of the parameter $\Phi_{0}^{D}$ turns out to be\[
0\leq\frac{\beta}{2}\Phi_{0}^{D}\equiv\varphi_{0}\leq\frac{\pi}{2}\]
 Static solutions of the equation of motions satisfying the appropriate
boundary conditions can be obtained by integrating the static equation
of motions: \begin{equation}
\partial_{x}^{2}\Phi=\frac{m^{2}}{\beta}\sin\beta\Phi\quad\longrightarrow\quad\frac{1}{2}(\partial_{x}\Phi)^{2}=\hat{C}+\frac{m^{2}}{\beta^{2}}(1-\cos\beta\Phi)\label{elsint}\end{equation}
 Introducing a {}``time'' variable $t=-mx$ and an angular coordinate
$q=\beta\Phi$, this system describes a pendulum of total energy $\hat{C}$.
The energy of a static solution is given by \[
E[\Phi(x)]=\int_{-\infty}^{0}dx\left(\frac{1}{2}\left(\partial_{x}\Phi\right)^{2}+V\left(\Phi(x)\right)\right)+V_{B}\left(\Phi(0)\right)\]
 and the requirement of having finite energy solutions in the field
theory fixes the boundary condition $q(t=\infty)=2\pi n$, i.e. $\hat{C}=0$.
This system is identical to a classical particle of unit mass moving
in the potential $\cos q-1$ as shown on the figure:

\begin{center}\includegraphics[%
  height=4cm]{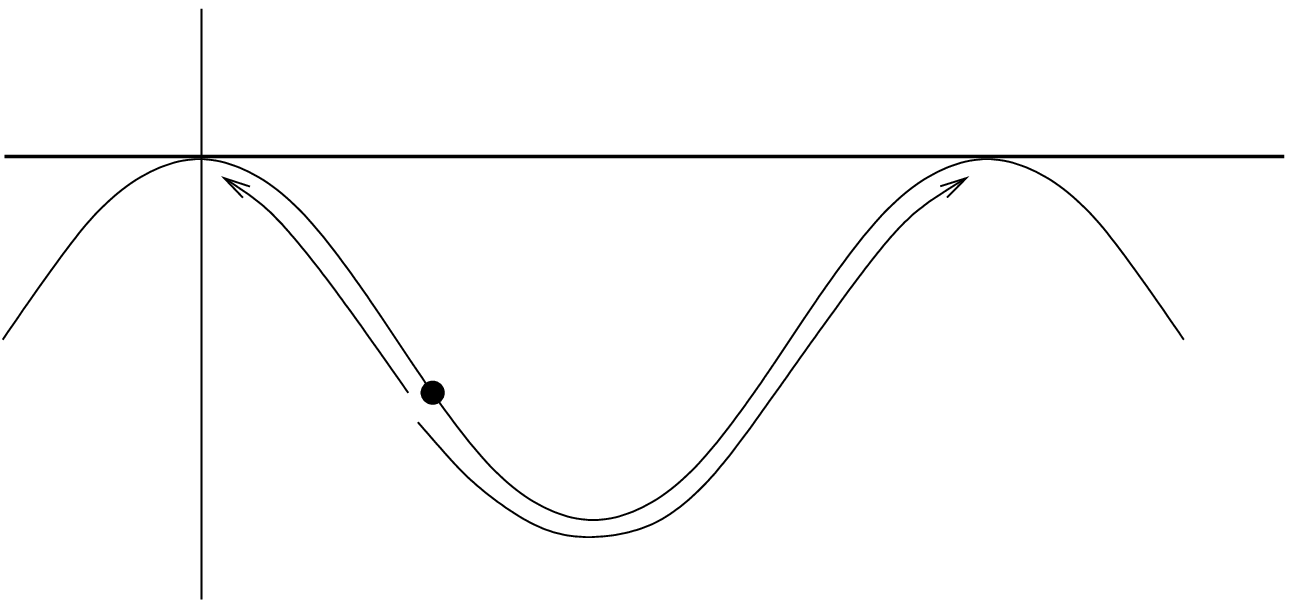}\end{center}

The boundary condition gives the starting position $q(t=0)=2\varphi_{0}$.
There are two solutions satisfying the boundary conditions which describe
a particle arriving at one of the two nearest potential maxima in
infinite time.

The sine-Gordon analogue of this picture is a soliton (+) or an anti-soliton
(-) standing at a particular position:

\begin{center}\includegraphics[%
  width=7cm]{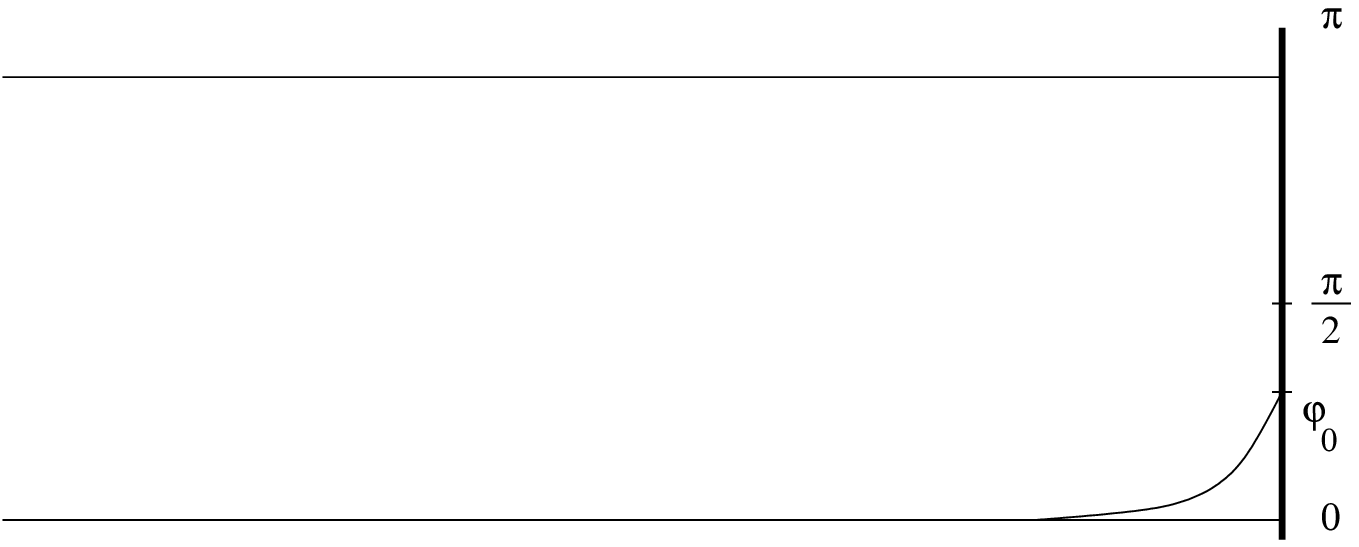}~~~\includegraphics[%
  width=7cm]{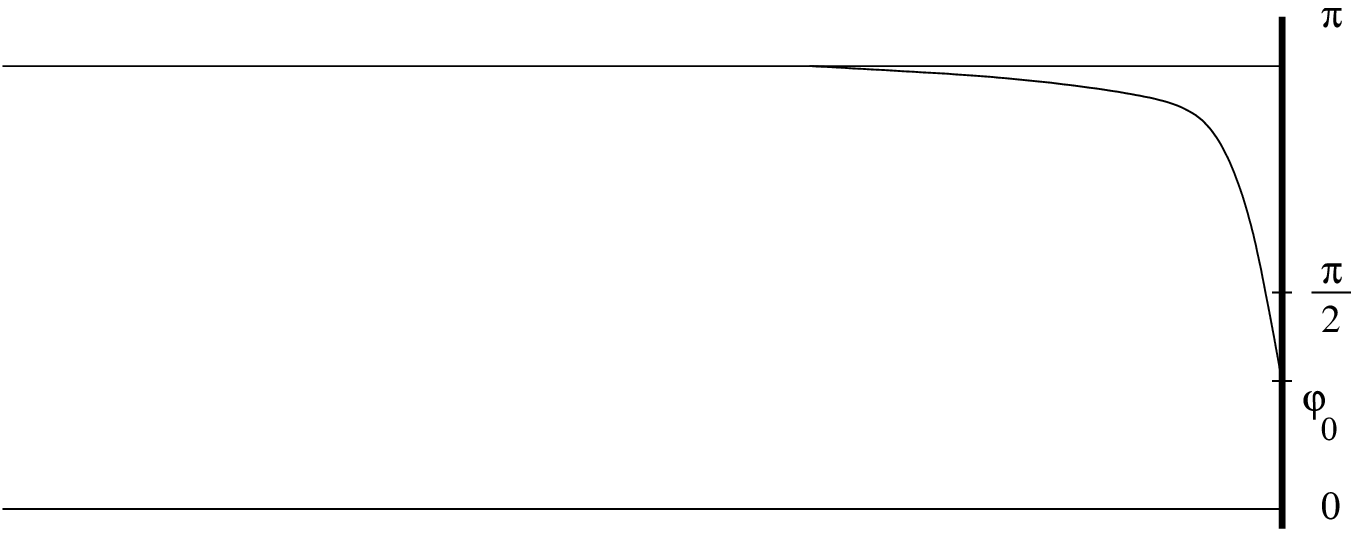}\end{center}

\[
\Phi_{+}\left(x,a^{+}\right)=\frac{4}{\beta}\arctan\left(e^{m(x-a^{+})}\right)\quad;\quad\Phi_{-}\left(x,a^{-}\right)=\frac{2\pi}{\beta}-\frac{4}{\beta}\arctan\left(e^{m(x-a^{-})}\right)\]
 where $\sinh(ma^{\pm})=\pm\cot(\varphi_{0}).$ The energies of these
two solutions are \begin{equation}
E_{\pm}(\varphi_{0})=\frac{4m}{\beta^{2}}(1\mp\cos\varphi_{0})\label{eq:eplusminus}\end{equation}
 and they approach their asymptotics at $x\rightarrow-\infty$ exponentially
fast:\begin{equation}
\Phi_{+}\left(x,a^{+}\right)=O\left(\mathrm{e}^{-m\left|x\right|}\right)\quad,\quad\Phi_{-}\left(x,a^{-}\right)=\frac{2\pi}{\beta}+O\left(\mathrm{e}^{-m\left|x\right|}\right)\quad\mathrm{as}\quad x\rightarrow-\infty.\label{eq:semiinf}\end{equation}

The quantum corrections to these energies can be obtained by analyzing
the small fluctuations around the appropriate backgrounds: \[
\left[-\frac{d^{2}}{dx^{2}}+m^{2}-\frac{2m^{2}}{\cosh^{2}(m[x-a^{\pm}])}\right]\xi_{\pm}(x)=\omega^{2}\xi_{\pm}(x)\quad;\quad\xi_{\pm}(0)=0\quad.\]
The solutions consist of a discrete and a continuous spectrum. 

Introducing $\epsilon\geq0$ as $\omega^{2}=m^{2}(1-\epsilon^{2})$
the normalizable modes of the \emph{discrete spectrum} are \[
\xi_{\pm}(x)\propto e^{m\epsilon(x-a^{\pm})}(\epsilon-\tanh[m(x-a^{\pm})]),\]
 where from the boundary condition we have $\epsilon=\mp\cos\varphi_{0}$.
Clearly we have a discrete mode only for the anti-solitonic ground
state which in the light of \cite{KP} corresponds to a boundary excited
state.

The \emph{continuous spectrum} can be described in terms of $q\geq0$
($\omega^{2}=m^{2}+q^{2})$ as \[
\xi_{\pm}(x)=\tilde{A}_{\pm}e^{-iq(x-a^{\pm})}\frac{iq+m\tanh(m[x-a^{\pm}])}{iq+m}+\tilde{B}_{\pm}e^{iq(x-a^{\pm})}\frac{iq-m\tanh(m[x-a^{\pm}])}{iq-m}\]
 and the ratios $\tilde{A}_{\pm}/\tilde{B}_{\pm}$ are determined
by the boundary conditions at $x=0$. The asymptotic form of the fluctuations
$(x\rightarrow-\infty)$ has the following form:\begin{equation}
\xi_{\pm}(x)\rightarrow C_{\pm}(e^{iqx}+e^{-iqx}e^{i\delta^{\pm}(q)})\label{hlrefb}\end{equation}
 It has an interpretation that the elementary excitation, the linearized
form of the first breather, reflects back from the boundary with classical
reflection coefficient: \[
e^{i\delta^{\pm}(q)}=\frac{m-iq}{m+iq}\,\frac{\pm\cos\varphi_{0}+\frac{iq}{m}}{\mp\cos\varphi_{0}+\frac{iq}{m}}\qquad.\]
Observe that the discrete mode can be obtained from the boundary condition
dependent pole term of the reflection factor and it has an interpretation
as the first breather bound to the boundary. The other pole singularity
is of kinematical origin. 

The semiclassical correction to the energy levels can be obtained
by summing up the frequencies of these fluctuations. By performing
the appropriate ultraviolet regularization \cite{KP} at the same
time we arrive at the final answer:\[
E_{+}^{\mathrm{semiclass}}(\varphi_{0})-E_{-}^{\mathrm{semiclass}}(\varphi_{0})=m\left(\frac{8}{\beta^{2}}-\frac{1}{\pi}\right)\cos\varphi_{0}+m\left(\frac{1}{2}-\frac{\varphi_{0}}{\pi}\right)\sin\varphi_{0}\]
 These result, namely the energy of the excited boundary state, the
reflection factor of the first breather and the energy differences
are in complete agreement with the classical limit of the quantum
results, given in appendix A, if we make the identification $\varphi_{0}=\eta_{\mathrm{cl}}$
between the parameters.

The system defined on the positive half line can be obtained from
the previous one by the $\mathcal{P}:\, x\rightarrow-x$ parity transformation.
See \cite{BG} for the details of how the two systems are related.
Clearly the roles of the solitons and anti-solitons are exchanged.

\section{Finite volume: general considerations}

\subsection{Static (vacuum) solutions}

Now we consider sine-Gordon theory restricted onto a strip with two
integrable boundary conditions. The action in the general case has
the form\[
\mathcal{A}=\int_{-\infty}^{\infty}dt\left[\int_{0}^{L}dx\:\mathcal{L}_{SG}(x,t)+V_{B}^{(0)}\left(\Phi(0,t)\right)-V_{B}^{(L)}\left(\Phi(L,t)\right)-\alpha\frac{\partial\Phi(L,t)}{\partial t}\right]\]
 The last term (proportional to $\alpha$) is allowed by integrability,
and on the strip - in contrast to the theory on the half line - it
cannot be eliminated from the action by an appropriate {}``gauge''
transformation \cite{AN}. Since this term is a total time derivative,
it gives no contribution to the classical equations of motion, and
we omit it in the sequel.

The equations of motion and the boundary condition are\begin{eqnarray*}
 &  & \partial_{t}^{2}\Phi-\partial_{x}^{2}\Phi+\frac{m^{2}}{\beta}\sin\beta\Phi=0\\
 &  & \partial_{x}\Phi|_{x=z}=-\frac{\beta M_{z}}{2}\sin\frac{\beta}{2}\left(\Phi(z,t)-\Phi_{z}\right)\qquad,\qquad z=0,\, L\end{eqnarray*}
 Dirichlet boundary conditions again can be obtained as limits in
which $M_{z}\rightarrow\infty$ at the appropriate end of the strip.

Ground states (vacua) of the model on a strip are static solutions
of the equations of motion, i.e. $\partial_{t}\Phi=0$. It can be
easily verified that such solutions provide extrema of the static
energy functional\begin{equation}
E[\Phi(x)]=\int_{0}^{L}dx\left(\frac{1}{2}\left(\partial_{x}\Phi\right)^{2}+V\left(\Phi(x)\right)\right)-V_{B}^{(0)}\left(\Phi(0)\right)+V_{B}^{(L)}\left(\Phi(L)\right)\label{eq:staticenergyfunctional}\end{equation}
 Using the first integral (\ref{elsint}), the solution, $\phi(x),$
can be written in the general form\begin{equation}
x=\int_{\Phi(0)}^{\phi(x)}\frac{du}{\pm\sqrt{2\left(V\left(u\right)+\hat{C}\right)}}\label{eq:genstatic}\end{equation}
 The sign ambiguity must be fixed from a more detailed analysis which
is discussed later. The two integration constants $\Phi(0)$ (the
value of the field at $x=0$) and $\hat{C}$ are fixed by the boundary
conditions and the volume. This integral can be written in terms of
Weierstrass $\mathfrak{p}$-function (see explicit examples later).

It is obvious that $E[\Phi(x)]$ is bounded from below. In addition,
the potential terms are also bounded and in fact periodic under $\Phi(x)\rightarrow\Phi(x)+4\pi/\beta$.
These properties guarantee that there always exists at least one static
solution, but we shall see that in many cases there are in fact two
different ones (solutions related by a period shift will be considered
the same).

Let us consider the limit $L\rightarrow\infty$. In that case finite
energy requires that \[
\Phi(x)\rightarrow\frac{2\pi n}{\beta}\]
 for any $x$ far away from the endpoints $x=0$ and $x=L$. Due to
(\ref{eq:semiinf}), we can approximate the solution by sewing together
two vacuum solutions, corresponding to imposing the boundary conditions
at $x=0$ ($x=L$) on a semi-infinite system on the half line. In
this way we get the finite volume solution up to a precision of $\exp(-mL)$,
(higher precision will be described later on), but there is a condition
that the asymptotic values of the left and right solutions must match
each other. The matching condition leads to some selection rules:
not all possible pairings of boundary states can be realized on a
strip. The selection rules can be described as follows. The general
boundary excited states can be characterized by giving a string of
integer numbers\[
\left|n_{1},n_{2},\dots n_{k}\right\rangle \]
(these are the boundary states corresponding to the poles $i\nu_{n_{1}},iw_{n_{2}},\dots$in
the notation used in \cite{genpap,DM}). The two static solutions
in infinite volume correspond to the ground state $\left|\right\rangle $
and first excited state $\left|0\right\rangle $. 

Let us first treat the case of Dirichlet boundary conditions. In \cite{DM}
the authors introduce the parity of a boundary state as the parity
of the number of integer labels, i.e. of $k$. For the two static
solutions, $\left|\right\rangle $ has even parity and the asymptotics
of the solution is \begin{equation}
\Phi(x)\rightarrow0\qquad\mathrm{as}\qquad x\rightarrow-\infty\label{eq:evenasypmtotics}\end{equation}
and $\left|0\right\rangle $ has odd parity with asymptotics\begin{equation}
\Phi(x)\rightarrow\frac{2\pi}{\beta}\qquad\mathrm{as}\qquad x\rightarrow-\infty\label{eq:oddasymptotics}\end{equation}
 This can be extended to the general case since the boundary states
$\left|n_{1},n_{2},\dots n_{k}\right\rangle $ can be obtained by
binding solitons/antisolitons to the boundary in an alternating sequence.
So in general even states correspond to configurations with asymptotics
(\ref{eq:evenasypmtotics}), while odd ones to (\ref{eq:oddasymptotics}).
Since the configuration on the strip must be continuous, we can infer
that for states containing no particle the only possible combinations
of left and right boundary states are either even-even or odd-odd.
This is exactly what was seen using truncated conformal space approach
in \cite{BPT2}. 

If we look at states containing particles, then an obvious extension
of the above argument yields that states containing even number of
solitonic particles obey the same matching condition for the boundary
states, while those with odd number of solitonic particles obey the
reverse (i.e. the boundary states at the two ends must have opposite
parity for these cases). This is meaningful because topological charge
is a conserved quantum number for Dirichlet boundary conditions.

The same selection rules can be extended to any boundary conditions,
which is simple to understand from the fact that the Dirichlet boundary
condition can be obtained by taking an appropriate limit of the general
one. This limit is given in terms of the Ghoshal-Zamolodchikov \cite{GZ}
boundary parameters $\eta$ and $\vartheta$ by taking $\vartheta\rightarrow\infty$,
while the boundary state spectrum only depends on $\eta$ \cite{genpap}.
As a result, it is natural to expect that selection rules on the strip
must be left unchanged by {}``undoing'' this limiting procedure.
The rule concerning states containing solitonic excitations can be
meaningfully transferred to the general case ($\vartheta<\infty$)
because topological charge modulo $2$ is conserved for general boundary
conditions as well.

\subsection{Small oscillations}

We can now investigate small oscillations around a static solution.
Putting\[
\Phi(x,t)=\phi(x)+\xi(x)\exp(-i\omega t)\]
 (where $\phi(x)$ is a static solution) and linearizing in $ \xi(x)$
we obtain the following equations of motion:\begin{eqnarray}
 &  & -\frac{d^{2}\xi}{dx^{2}}+V''\left(\phi(x)\right)\xi(x)=\omega^{2}\xi(x)\nonumber \\
 &  & \partial_{x}\xi(x)|_{x=z}=M_{z}\frac{\beta^{2}}{4}\cos\frac{\beta}{2}\left(\phi(z)-\Phi_{z}\right)\xi(z)\quad,\quad z=0,\, L\label{genstab}\end{eqnarray}
 For Dirichlet boundary condition at one (or both) ends we get that
$\xi(x)$ must vanish at the appropriate boundary point(s). Inserting
the general static solution (\ref{eq:genstatic}) one gets Lame's
equation of order $1$, with linear boundary conditions for $\xi(x)$.

\begin{figure}
\begin{center}\includegraphics{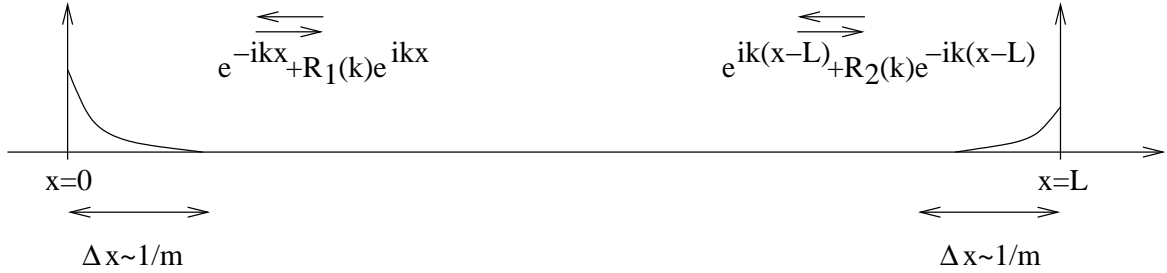}\end{center}

\caption{\label{bethe-yang} Derivation of the classical Bethe-Yang equation}
\end{figure}

For the moment let us only consider the limit $L\rightarrow\infty$.
Since in this limit the static background is exponentially close to
a superposition of two solitons localized at the boundary, we can
apply the asymptotic solution of the semi-infinite case at any point
$x$ sufficiently far from the boundaries (see Fig. \ref{bethe-yang}).
This means that the solution can be written in two different asymptotic
forms:\begin{eqnarray*}
 &  & \xi(x)\,\sim\, A\left(\mathrm{e}^{-ikx}+R_{1}(k)\mathrm{e}^{ikx}\right)\quad\textrm{as $x$ $\textrm{is large}$}\\
 &  & \xi(x)\,\sim\, B\left(\mathrm{e}^{ik(x-L)}+R_{2}(k)\mathrm{e}^{-ik(x-L)}\right)\quad\textrm{as $x-L$ $\textrm{is large}$}\end{eqnarray*}
 where $R_{1}(k)$ and $R_{2}(k)$ are the appropriate reflection
factors calculated on the (semi-infinite) half line (\ref{hlrefb}).
The two asymptotic forms must match in any point $x$ in the middle%
\footnote{Using standard quantum mechanics, the wave function $\xi(x)$ and
its derivative $\xi'(x)$ must be continuous, which translates to
the exact matching of the two asymptotic forms for any value of $x$.%
}, which gives two homogeneous linear equations for $A$ and $B$.
The compatibility of these equations requires that \[
R_{1}(k)R_{2}(k)\mathrm{e}^{2ikL}=1\]
 This is the correct classical limit of the Bethe-Yang equations for
a state containing a single first breather (i.e. the fundamental particle)
in sine-Gordon theory on a strip. This general argument shows that
the analysis of the small oscillations around a static solutions must
reproduce the correct semiclassical limit of the state containing
a breather between the corresponding boundary states. Later we show
this explicitly for a number of cases as a consistency check of our
analysis.

We remark that one can similarly derive that for periodic boundary
condition and a background containing a single soliton in finite volume
$L$, the appropriate equation is\[
\mathrm{e}^{ikL}S_{SB_{1}}(k)=1\]
 where $S_{SB_{1}}(k)$ is the classical limit of the phase shift
sustained by $B_{1}$ when scattered by a soliton. The derivation
in this case proceeds by matching the two asymptotics of the scattered
wave in a point of the finite volume far away from the location of
the soliton. It is straightforward to check that this relation is
a consequence of the formulae describing the small oscillations around
a solitonic background in \cite{MRS}.

\section{Sine-Gordon model on the strip: Dirichlet-Dirichlet boundaries }

We start by considering the most simple type of boundary conditions
which preserves integrability, namely Dirichlet boundary conditions
on both end: the value of the field at the origin and at $L$ are
fixed to be \[
\Phi(0)=\Phi_{0}^{D}\quad,\quad\Phi(L)=\Phi_{L}^{D}\]
 Using the various bulk symmetries of the model $\Phi_{0}^{D}$ can
be mapped into the fundamental range\[
0\leq\frac{\beta}{2}\Phi_{0}^{D}\equiv\varphi_{0}\leq\frac{\pi}{2}\quad.\]
 If we would like to describe all the solitonic sectors of this theory
we have to consider the various boundary conditions \[
\Phi(L)=\Phi_{L}^{D}+\frac{2\pi}{\beta}n\quad n\in Z\]
 simultaneously, thus we will suppose that $0\leq\Phi_{L}^{D}<\frac{2\pi}{\beta}$.

\subsection{The case $\Phi_{L}^{D}=0$}

As a starting point we consider the case when $\Phi_{L}^{D}=0.$

For any $n$ we have one and only one static ground state, which can
be obtained by integrating the first integral of the static equation
of motion (\ref{elsint}). In the language of the classical particle
it means that the particle has to reach the top of the hill separating
the sectors in finite time $L$ as shown on the figure:

\begin{center}\includegraphics[%
  height=4cm]{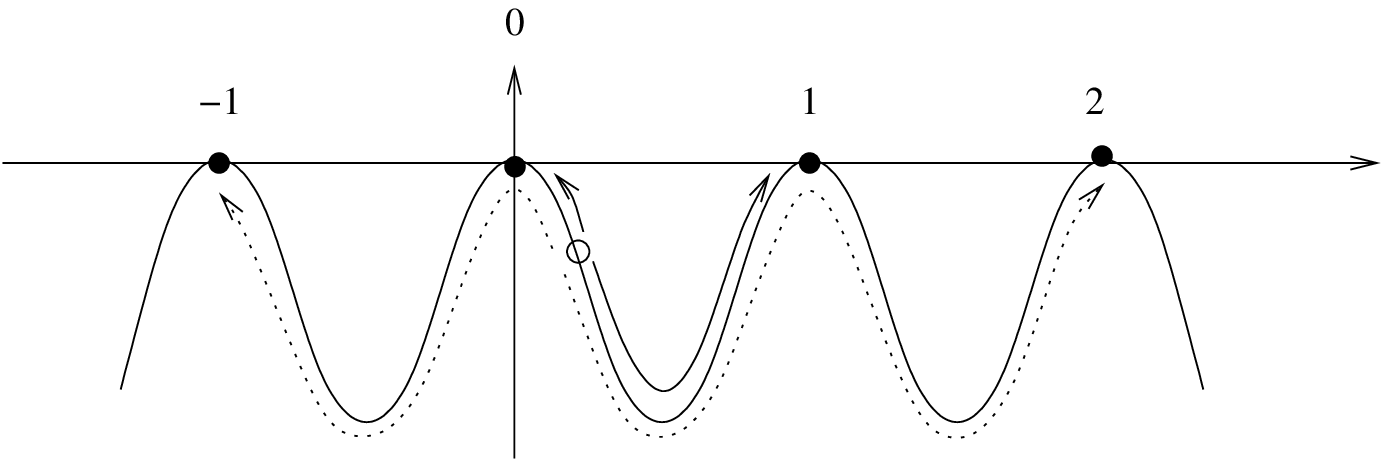}\end{center}

Clearly for this reason in any sector the energy of the classical
particle is positive (and consequently $\hat{C}>0$). For the sectors
$0,-1,\dots$ the velocity is negative while for $1,2\dots$it is
positive. The actual form of the solution, $\phi(x)$ in the $\phi(L)=0$
(+) and $\phi(L)=\frac{2\pi}{\beta}$ (-) sectors are

\begin{center}\includegraphics[%
  width=7cm]{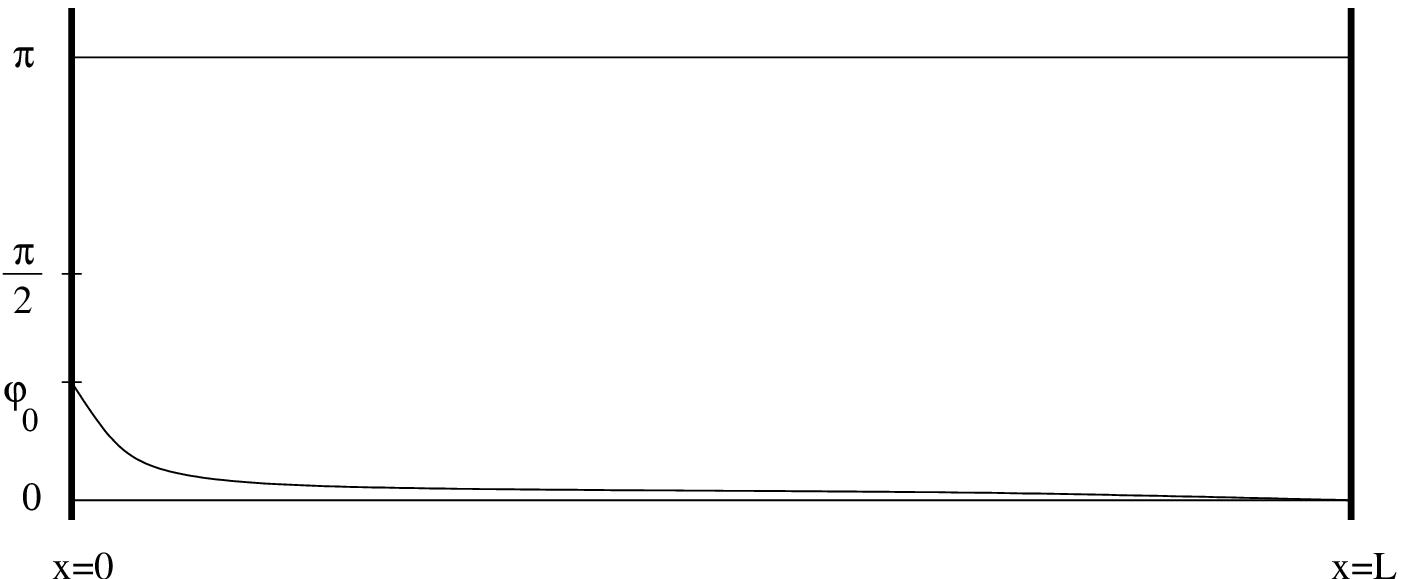}~~~\includegraphics[%
  width=7cm]{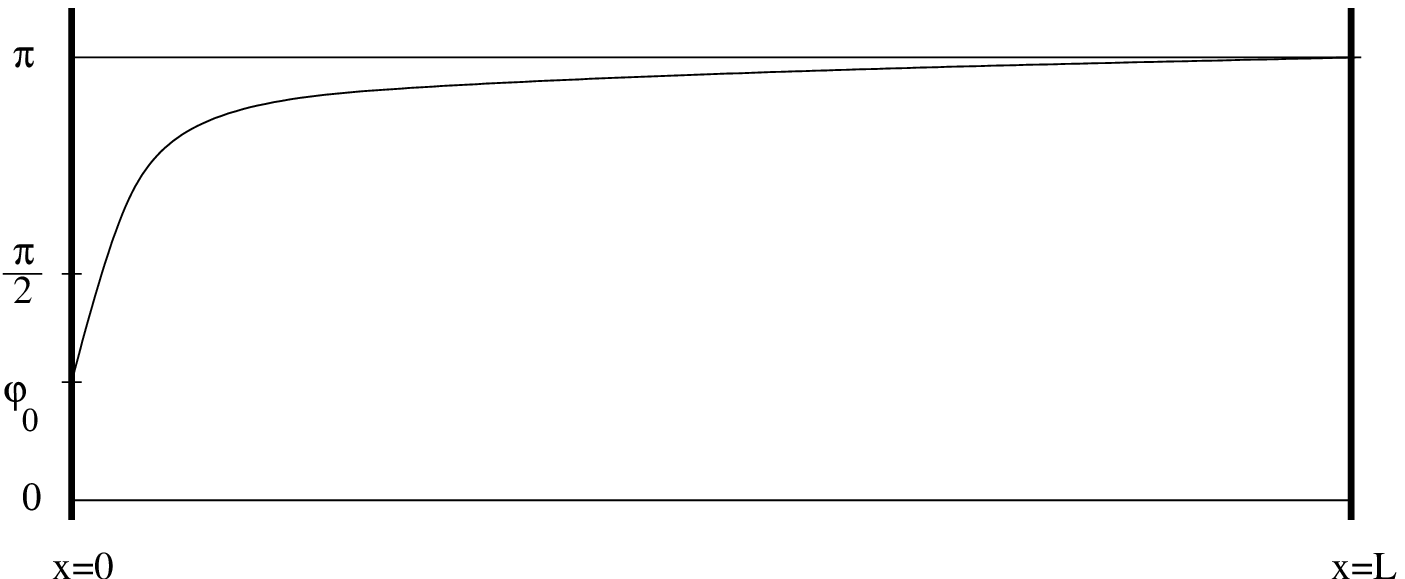}\end{center}

\[
mx=\int_{\beta\phi/2}^{\varphi_{0}}\frac{du}{\sqrt{\sin^{2}u+C}}\quad(+),\quad mx=\int_{\varphi_{0}}^{\beta\phi/2}\frac{du}{\sqrt{\sin^{2}u+C}}\quad(-)\quad\mathrm{where}\quad\hat{C}=\frac{2m^{2}}{\beta^{2}}C\,,\]
 and $C$ can be determined uniquely from the boundary conditions
and the width of the strip $l$ as \begin{equation}
l:=mL=\int_{0}^{\varphi_{0}}\frac{du}{\sqrt{\sin^{2}u+C}}\quad(+),\quad l=\int_{\varphi_{0}}^{\pi}\frac{du}{\sqrt{\sin^{2}u+C}}\quad(-)\:.\label{lasfC}\end{equation}
 Note that the solution $(+)$ is in the $n=0$, while the $(-)$
is in the $n=1$ topological sector. Once $C$ is known, the energies
of these solutions can be written as\begin{eqnarray}
E_{+}(\varphi_{0},l) & = & \frac{4m}{\beta^{2}}\left[\int_{0}^{\varphi_{0}}\sqrt{\sin^{2}u+C}du-\frac{Cl}{2}\right]\nonumber \\
E_{-}(\varphi_{0},l) & = & \frac{4m}{\beta^{2}}\left[\int_{\varphi_{0}}^{\pi}\sqrt{\sin^{2}u+C}du-\frac{Cl}{2}\right]\label{eq:energydir0}\end{eqnarray}
 In the large volume limit, $l\rightarrow\infty$, the solution has
the form of a standing anti-soliton (+) or soliton (-) with the corresponding
energies. As $0\leq\varphi_{0}\leq\frac{\pi}{2}$ the case (+) forms
the ground state, while the case (-) is an excited state. These solutions
can be obtained by parity transformation from the half-line results,
since the theory is defined on the positive half-line.

In order to analyze the small fluctuations around this background
we rewrite them in terms of Weierstrass's $\mathfrak{p}(z)$ function.
After a straightforward change of integration variables and an appropriate
deformation of the integration contour into the complex plane the
case (+) solution can be written as \[
xm=\int\limits _{P_{D}}^{\infty}\frac{dT}{\sqrt{4T^{3}-g_{2}T-g_{3}}}-\int\limits _{P_{\phi}}^{\infty}\frac{dT}{\sqrt{4T^{3}-g_{2}T-g_{3}}},\]
 where \[
P_{D}=\frac{1-C}{3}-\sin^{2}\varphi_{0};\qquad P_{\phi}=\frac{1-C}{3}-\sin^{2}\frac{\beta\phi}{2},\]
 and \[
g_{2}=\frac{4}{3}(1+C+C^{2}),\qquad g_{3}=-\frac{4}{27}(1-C)(1+2C)(2+C).\]
 Since the integrals appearing here are the standard integral representations
of $\mathfrak{p}(z)$ \cite{whit} \cite{riz} we can write \begin{equation}
P_{\phi}=\frac{1-C}{3}-\sin^{2}\frac{\beta\phi(x)}{2}=\mathfrak{p}(d-mx)\:,\quad P_{D}=\mathfrak{p}(d)\,.\label{Wei1}\end{equation}
 where\begin{equation}
d=\int\limits _{P_{D}}^{\infty}\frac{dT}{\sqrt{4T^{3}-g_{2}T-g_{3}}}\label{wei2}\end{equation}
 The solution, $\phi(x)$ in the other case (-) is simply \[
P_{\phi}=\frac{1-C}{3}-\sin^{2}\frac{\beta\phi(x)}{2}=\mathfrak{p}(d+mx)\]
 Since $C>0$ the roots of the cubic polynomial defining ${\mathfrak{p}}(z)$
are \begin{equation}
e_{3}=\frac{-C-2}{3}<e_{2}=\frac{1-C}{3}<e_{1}=\frac{1+2C}{3},\label{ckp}\end{equation}
 thus the half periods of ${\mathfrak{p}}(z)$ are \cite{riz}\begin{equation}
\omega_{1}=\frac{K(k)}{\sqrt{1+C}},\quad\omega_{2}=\frac{iK(k^{\prime})}{\sqrt{1+C}},\quad k=\frac{1}{\sqrt{1+C}},\quad k^{\prime}=\sqrt{\frac{C}{1+C}}.\label{ckpp}\end{equation}
 Using the well known relation between ${\mathfrak{p}}(z)$ and Jacobi's
elliptic functions \cite{riz} the classical solution $\phi(x)$ can
be given explicitly. One finds from (\ref{Wei1}) in terms of the
dimensionless $xm\mapsto x$ variable\begin{equation}
\frac{\beta\phi(x)}{2}=\frac{1}{i}\ln\left(\frac{\sqrt{1+C}\left(\mathrm{dn}(d\mp x,k)\pm1\right)}{\mathrm{sn}(\sqrt{1+C}\left(d\mp x\right),k)}\right)\;\qquad.\label{masodik}\end{equation}
 where $d$ is determined from \[
\cos^{2}\varphi_{0}=\frac{1+C}{{\textrm{sn}}^{2}(d\sqrt{1+C},k)}\]

\subsubsection{Infinite volume limit: leading corrections}

It is interesting to consider the $l\,\rightarrow\,\infty$ limits
of the classical ground state energies, since they are proportional
to the classical limit of the scaling functions ({}``effective central
charge''). Using (\ref{lasfC}) and the asymptotic expansion of elliptic
integrals, we obtain the asymptotic value of $C$:\[
\sqrt{C}\sim4 \tan\frac{\varphi_{0}}{2}\,\mathrm{e}^{-l}\qquad,\qquad l\rightarrow\infty\,.\]
Plugging this into the formulae for $E_{\pm}$ (\ref{eq:energydir0}),
after some effort we obtain \begin{eqnarray}
E_{+}\left(\varphi_{0},l\right) & \sim & \frac{4m}{ \beta^{2}}\left(1-\cos\varphi_{0}+4 \tan^{2}\frac{\varphi_{0}}{2}\,\mathrm{e}^{-2l}\right)\nonumber \\
E_{-}\left(\varphi_{0},l\right) & \sim & \frac{4m}{ \beta^{2}}\left(1+\cos\varphi_{0}+4 \cot^{2}\frac{\varphi_{0}}{2}\,\mathrm{e}^{-2l}\right)\label{eq:energyasdir0}\end{eqnarray}
It is a very interesting challenge to find a quantum field theoretic
derivation of these formulae following the general description of
finite size effects originally put forward by Lüscher for periodic
boundary conditions \cite{Lusch}. This expectation is supported by
the results of \cite{MRS}, where, in the periodic case a full agreement
is found between Lüscher's results and the infrared expansion ($l\rightarrow\infty$
limit) of the kink's classical energy.

\subsubsection{Stability analysis}

In the semiclassical analysis the field is expanded, $\Phi(x,t)=\phi(x)+\xi(x)e^{-i\omega t}$,
around the solution of the static equation of motion, $\phi(x)$.
The small oscillations are investigated in the approximation when
both the equation of motion and the boundary condition are linearized.
Introducing the dimensionless variable $mx\mapsto x$ the boundary
conditions and the equation of motion (\ref{genstab}) take the following
form: \begin{equation}
\frac{d^{2}\xi}{dx^{2}}+2\sin^{2}\left(\frac{\beta\phi(x)}{2}\right)\xi+p^{2}\xi=0\quad,\quad\xi(0)=\xi(l)=0,\label{stabeq}\end{equation}
 where $\omega^{2}=m^{2}(1+p^{2})$. Using the representation in (\ref{Wei1}),
this can be rewritten as \begin{equation}
\frac{d^{2}\xi}{dx^{2}}-[2\mathfrak{p}(d\mp x)+a]\xi(x)=0,\qquad a=-\frac{2}{3}(1-C)-p^{2},\label{lame}\end{equation}
 which is the standard form of Lame's equation with $n=1$ (cf. \cite{OP}).
The general solution of this equation is known \cite{kamk}; if we
introduce $y=d\mp x$ and $\xi(x)\rightarrow\nu(y)$ then it is given
by \begin{equation}
\nu(y)=\frac{1}{\sigma(y)}[A\sigma(y+\alpha)e^{-y\zeta(\alpha)}+B\sigma(y-\alpha)e^{y\zeta(\alpha)}],\label{gensol}\end{equation}
 where $A$, $B$ are constants, $\alpha$ is the solution of $\mathfrak{p}(\alpha)=a$
and \[
\zeta(u)=\frac{1}{u}-\int\limits _{0}^{u}\left(\mathfrak{p}(v)-\frac{1}{v^{2}}\right)dv,\qquad\sigma(u)=u\exp\left[\int\limits _{0}^{u}\left(\zeta(v)-\frac{1}{v}\right)dv\right]\]
 are Weierstrass's $\zeta$ and $\sigma$ functions. The two boundary
conditions on $\nu$ impose the following quantization condition on
$p$:\begin{equation}
-\frac{\sigma(d\mp l-\alpha)\sigma(d+\alpha)}{\sigma(d\mp l+\alpha)\sigma(d-\alpha)}\mathrm{e}^{\mp2l\zeta(\alpha)}+1=0\label{fulqud}\end{equation}
 The algorithm giving the quantized momenta from (\ref{fulqud}) is
the following: first $d$ and $C$ are determined from the boundary
parameter $\Phi_{0}^{D}$. (In practice it is simpler to treat $C$
as an input parameter and determine $d$ and $l$ from $C$). Then,
using these values in the \textsl{logarithmic version} of eqn. (\ref{fulqud}),
one determines $\alpha_{n}$ for each integer $n$ coming from the
substitution $1=\exp(2i\pi n)$. The $n$ dependent momenta are finally
given by \begin{equation}
p_{n}^{2}=-\frac{2}{3}(1-C)-\mathfrak{p}(\alpha_{n})\label{qmom}\end{equation}
 .

Equation (\ref{fulqud}) (or its logarithmic version) should be considered
as the (semi)classical version of the Bethe - Yang equation for the
first breather (fundamental scalar particle) in finite volume $l$.
To support this interpretation we consider the $l\rightarrow\infty$
limit, which we implement by setting $C$ to zero. Then $e_{1}=e_{2}=1/3$,
$e_{3}=-2/3$, and $\mathfrak{p}$ simplifies \cite{riz} \[
\mathfrak{p}(u)=-\frac{2}{3}+\coth^{2}(u).\]
 Therefore the equations determining $\alpha$ and $d$ become\begin{equation}
\coth\alpha=ip\;,\qquad\coth^{2}d=\cos^{2}\varphi_{0}\label{aseq}\end{equation}
 Furthermore in this case $\zeta$ and $\sigma$ can be written explicitly
as \[
\zeta(u)=-\frac{u}{3}+\coth(u),\qquad\sigma(u)=\sinh(u)\exp[-\frac{u^{2}}{6}].\]
 In the asymptotically large $l$ domain one finds from (\ref{fulqud})\begin{equation}
-\frac{ip-1}{ip+1}\frac{ip\pm\cos\varphi_{0}}{ip\mp\cos\varphi_{0}}\mathrm{e}^{2ilp}+1=0\label{ntaszd}\end{equation}
 To obtain this we wrote $d=d_{0}+i\frac{\pi}{2}$ in accord with
(\ref{aseq}), where $\tanh(d_{0})=-\cos\varphi_{0}$, since the asymptotic
background is an antisoliton rather than a soliton standing in the
vicinity of the $x=0$ boundary. In fact we verified numerically that
making the substitution $T=P_{D}+iy$ in the integral (\ref{wei2})
yields \[
d=\int\limits _{0}^{\infty}\frac{idy}{\sqrt{4(P_{D}+iy)^{3}-g_{2}(P_{D}+iy)-g_{3}}}=\tilde{d}_{0}(C,\Phi_{0}^{D})+\omega_{2}(C),\]
 where $\tilde{d}_{0}$ is real and the purely imaginary $\omega_{2}(C)$
is equal - up to our numerical precision - to the second half period
of ${\mathfrak{p}}$, which, for $C\rightarrow0$ indeed $\omega_{2}(C)\rightarrow i\frac{\pi}{2}$.
In the same limit it was also found that $\tilde{d}_{0}\rightarrow d_{0}$
with $\tanh(d_{0})=-\cos\varphi_{0}$.

As shown in appendix A the functions multiplying $e^{2ilp}$ in eqn.
(\ref{ntaszd}) are the (semi)classical limits of the first breather's
reflection amplitudes on various Dirichlet boundaries with parameter
$\varphi_{0}$ in infinite half space. This confirms the interpretation
of eqn. (\ref{fulqud}) as the classical limit of the finite width
Bethe-Yang equation.

The essential difference between the two cases in eqn. (\ref{fulqud},
\ref{ntaszd}) is that for the lower sign - that corresponds to the
background describing the excited ground state - the function multiplying
$e^{2l\zeta(\alpha)}$ ($e^{2ilp}$) admits a pole, that for $l\rightarrow\infty$
may be interpreted as a bound state, while no such pole is present
for the upper sign - that corresponds to the ground state background.

In the half line case poles of the reflection functions provide information
on the discrete spectrum, which corresponds to excited boundary states.
Now the potential poles come from the quantization conditions (\ref{fulqud}). 

Certain poles of the functions multiplying $e^{2l\zeta(\alpha)}$
in the various quantization conditions may correspond to bound states
of the finite width problem. Other poles of the same functions may
describe some kinematical singularities. The difference between these
two types of poles is in the way they depend on the boundary parameters
and on the width of the strip: the potential bound state poles depend
on both the boundary parameters and on $l$, while the kinematical
poles are independent of them. Furthermore the (semi)classical energy
of the bound state in finite strip \begin{equation}
\omega=m\sqrt{1+p^{2}}=m\sqrt{1-\frac{2(1-C)}{3}-\mathfrak{p}(\alpha)}\label{bsp}\end{equation}
 should go, for large $l$-s, to the classical limit of the energy
of the known bound state in the half line theory.

Consider the quantization condition (\ref{fulqud}). With appropriate
signs in the arguments the function appearing here has a potential
bound state pole and the energy of the corresponding bound state is
\begin{equation}
\omega^{D}=m\sqrt{C+\sin^{2}\varphi_{0}}.\label{dene}\end{equation}
 Note that for $C\rightarrow0$ ($l\rightarrow\infty$) $\omega^{D}\rightarrow m\sin\varphi_{0}$
indeed. The bound state is present only for the excited, solitonic,
background, which is consistent with the $l\rightarrow\infty$ asymptotic.

\subsection{The general Dirichlet case}

Now we relax the condition $\Phi_{L}^{D}=0$ and describe the general
case. Keeping all the possible sectors of the theory we allow the
$\Phi(L)=\Phi_{L}^{D}+\frac{2\pi n}{\beta}$ boundary conditions.
For any $n$ we have a sector of the theory. The static ground states
can be visualized in the language of the classical particle as shown
on the figure

\begin{center}\includegraphics[%
  height=4cm]{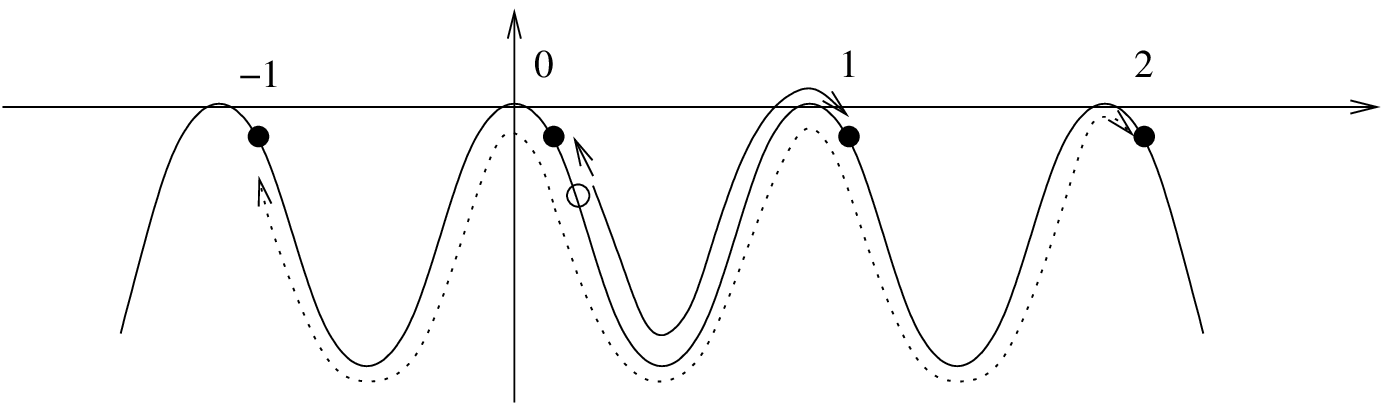}\end{center}

For the sectors labeled by $-1,-2,\dots$ there exists a unique static
solution. The energy of the particle is positive while the velocity
is negative. For the sector $1,2,\dots$ we have the same statement:
the static solution is unique, the energy of the particle is also
positive just as is its velocity. The $n=0$ sector is the most complicated
one. Here the energy is negative but we do not have a unique static
solution. The two lowest lying static solutions correspond to the
following motions of the classical particle:

\begin{center}\includegraphics[%
  height=4cm]{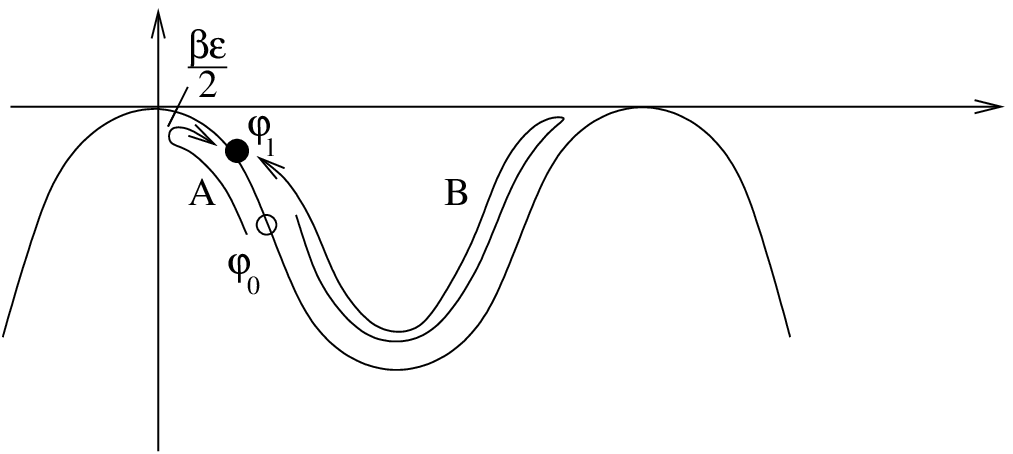}\end{center}

In the large volume limit $l\rightarrow\infty$ the energy of the
particle tends to zero from below and the solutions have interpretations
as two independent soliton and anti-soliton standing at the two ends
of the strip, i.e. we recover the solutions corresponding to two separate
Dirichlet boundary conditions corresponding to $\Phi_{0}^{D}$ and
$\Phi_{L}^{D}$. In the case (A) the solution has asymptotic energy
$E_{+}(\varphi_{0})+E_{+}(\varphi_{l})$ while for case (B) it is
$E_{-}(\varphi_{0})+E_{-}(\varphi_{l})$ (see eqn. (\ref{eq:eplusminus})
for notations). Since $0\leq\varphi_{0}\leq\frac{\pi}{2}$ the case
(A) solution has lower energy if $\varphi_{0}<\pi-\varphi_{l}$. We
can restrict ourselves to this case without loss of generality, since
the other possibility can be described by parity transforming the
strip and correspondingly inverting the time variable of the particle,
which exchanges the roles of the initial and final conditions. (Notice,
however, that increasing continously $\varphi_{l}$ over $\pi-\varphi_{0}$
the roles of the two solutions, namely the ground state and the first
excited states are exchanged). Now decreasing the volume the particle
has less and less energy; when the kinetic energy becomes exactly
zero at the one of the endpoints of the motion ($\varphi_{0}$ or
$\varphi_{l}$) then the (B) type solution ceases to exist and no
such solution exists for smaller volumes. At that point a breather
comes out from the boundaries and this time-dependent solution will
follow the energy line as it will be explained later. We can conclude
that although there are more static solutions in this sector for general
volume, only the one with the lowest energy survives in the small
volume limit.

The mechanism described above is clearly analogous to what was seen
before in \cite{BPT2,BPT1}, where in a different configuration an
excited boundary state could {}``decay'' in finite volume by emitting
a breather. The appropriate scaling function was described by the
Bethe-Yang equation of a single-particle state containing a first
breather with zero momentum quantum number below the critical volume,
and by continuing the rapidity of this particle to imaginary values
above the critical volume. The phenomenon described above is somewhat
different, since in this case the two boundaries can be thought of
as emitting a soliton and an antisoliton, respectively, which then
combine to form the breather inside the strip and it is not obvious
how to describe it in terms of some Bethe-Yang equation as was done
in \cite{BPT2,BPT1}.

\subsection{$n=0$ topological sector: non monotonic background}

\subsubsection{(A) type solution}

Now we consider the SG model on a strip with Dirichlet boundary conditions
at both ends with: \begin{equation}
0<\varphi_{0}<\frac{\pi}{2}\qquad,0\leq\varphi_{l}\leq\pi-\varphi_{0},\label{eq:fundamentalrange}\end{equation}
 and consider the case (A) solution first. In this case, for \textsl{sufficiently
large} $l$-s, the ground state is given qualitatively by the superposition
of a static anti-soliton and soliton located in the vicinity of the
two boundaries:

\begin{center}\includegraphics[%
  height=4cm]{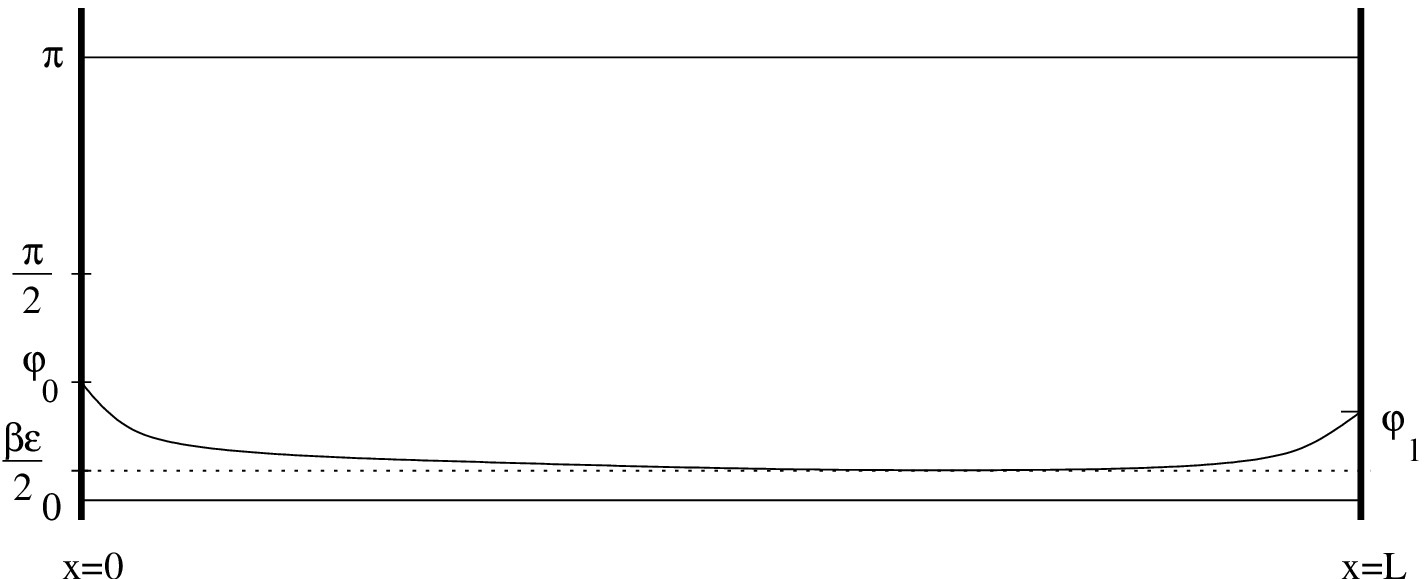}\end{center}

Therefore the solution first decreases from $\Phi_{0}^{D}$ to some
minimal value $\epsilon$ then increases from $\epsilon$ to $\Phi_{l}^{D}$
and in these two domains it is given by slightly different expressions.
Since $\partial_{x}\Phi$ must vanish at the turning point the integration
constant $C$ is \[
C=-\sin^{2}(\frac{\beta\epsilon}{2})\leq0.\]
 Using the dimensionless $x$ variable the solution can be written
as \[
x=\int\limits _{\beta\phi/2}^{\varphi_{0}}\frac{dv}{\sqrt{\sin^{2}v+C}},\qquad\epsilon\leq\phi<\Phi_{0}^{D},\]
 in the decreasing domain, while as \[
x=l_{-}+\int\limits _{\beta\epsilon/2}^{\beta\phi/2}\frac{dv}{\sqrt{\sin^{2}v+C}},\qquad\epsilon\leq\phi<\Phi_{l}^{D},\]
 in the increasing one. Here we introduced \[
l_{-}=\int\limits _{\beta\epsilon/2}^{\varphi_{0}}\frac{dv}{\sqrt{\sin^{2}v+C}},\]
 and the turning point is determined by the width of the strip as
\[
l=l_{-}+l_{+}=\int\limits _{\beta\epsilon/2}^{\varphi_{0}}\frac{dv}{\sqrt{\sin^{2}v+C}}+\int\limits _{\beta\epsilon/2}^{\varphi_{l}}\frac{dv}{\sqrt{\sin^{2}v+C}},.\]
 Note that $l\rightarrow\infty$ corresponds to $\epsilon\rightarrow0$.
The classical energy of the solution can be written as \[
E=\frac{4m}{\beta^{2}}\left(\frac{l}{2}\sin^{2}\frac{\beta\epsilon}{2}+\int\limits _{\beta\epsilon/2}^{\varphi_{0}}du\sqrt{\sin^{2}u+C}+\int\limits _{\beta\epsilon/2}^{\varphi_{l}}du\sqrt{\sin^{2}u+C}\right).\]
 The large $l$ limit of this energy has a simple interpretation as
the sum of the energies of the static solitons/antisolitons constituting
the ground state.

In both domains the solution can be expressed in terms of the Weierstrass's
$\mathfrak{p}(z)$. The difference now is that since $C\leq0$ the
roots of the cubic polynomial defining ${\mathfrak{p}}(z)$ are \begin{equation}
e_{3}=\frac{-C-2}{3}<e_{2}=\frac{1+2C}{3}<e_{1}=\frac{1-C}{3},\label{ckn}\end{equation}
 thus the half periods of ${\mathfrak{p}}(z)$ are \cite{riz}\begin{equation}
\omega_{1}=K(k),\quad\omega_{2}=iK(k^{\prime}),\quad k=\sqrt{1+C},\quad k^{\prime}=\sqrt{-C}.\label{cknp}\end{equation}
 In deforming the various integrals we also encounter the following
expression \[
I=\int\limits _{P_{\epsilon}}^{\infty}\frac{dT}{g(T)},\qquad P_{\epsilon}=\frac{1-C}{3}-\sin^{2}(\frac{\beta\epsilon}{2})=\frac{1+2C}{3}.\]
 Since $\frac{1+2C}{3}=e_{2}$, and ${\mathfrak{p}}(\omega_{1}+\omega_{2})=e_{2}$
thus $I=\omega_{1}+\omega_{2}$. Therefore, defining $d_{i}$ $i=0,l$
as previously \[
P_{D_{i}}\equiv\frac{1-C}{3}-\sin^{2}\varphi_{i}=\mathfrak{p}(d_{i}),\]
 one can write \[
l=d_{0}+d_{l}-2(\omega_{1}+\omega_{2}),\qquad l_{-}=d_{0}-(\omega_{1}+\omega_{2}),\quad l_{+}=d_{l}-(\omega_{1}+\omega_{2}),\]
 and the solution becomes \[
P_{\phi}=\mathfrak{p}(d_{0}-x),\]
 in the decreasing domain and \[
P_{\phi}=\mathfrak{p}(x-d_{0}+2(\omega_{1}+\omega_{2}))\]
 in the increasing one.

Using the relation between ${\mathfrak{p}}(z)$ and Jacobi's elliptic
functions \cite{riz} the classical solution $\phi(x)$ can be given
explicitly in this case also. One finds from (\ref{Wei1}) in terms
of the dimensionless $xm\mapsto x$ variable\begin{equation}
\frac{\beta\phi(x)}{2}=\frac{1}{i}\ln\left(\frac{1\pm\mathrm{cn}(d-x,k)}{\mathrm{sn}(d-x,k)}\right)\;.\label{elso}\end{equation}
 where $d$ is determined from \[
\cos^{2}\varphi_{0}=\frac{1}{\textrm{sn}^{2}(d,k)}\quad.\]

In the stability analysis the fluctuation equations are solved simultaneously
in the two domains by exploiting the $\mathfrak{p}(u)=\mathfrak{p}(-u)$
and $\mathfrak{p}(u+2(\omega_{1}+\omega_{2}))=\mathfrak{p}(u)$ invariances.
Using the coordinate $y=d_{0}-x$ the boundary condition at $x=0$
becomes \[
\nu(d_{0})=Af(d_{0},\alpha)+Bf(d_{0},-\alpha)=0,\qquad f(x,\alpha)=\frac{\sigma(x+\alpha)}{\sigma(x)}e^{-x\zeta(\alpha)}.,\]
 while at $x=l$ one finds \[
\nu(-d_{l}+2(\omega_{1}+\omega_{2}))=Af(-d_{l}+2(\omega_{1}+\omega_{2}),\alpha)+Bf(-d_{l}+2(\omega_{1}+\omega_{2}),-\alpha)=0.\]
 Non vanishing $A$, $B$ exist only if the \[
e^{2l\zeta(\alpha)}\frac{\sigma(d_{l}-\alpha-2(\omega_{1}+\omega_{2}))\sigma(d_{0}-\alpha)}{\sigma(d_{l}+\alpha-2(\omega_{1}+\omega_{2}))\sigma(d_{0}+\alpha)}=1\]
 condition holds. Using the quasi-periodicity of $\sigma(u)$ and
an interesting identity \cite{whit} one can write \begin{eqnarray}
\frac{\sigma(d_{l}-\alpha-2(\omega_{1}+\omega_{2}))\sigma(d_{0}-\alpha)}{\sigma(d_{l}+\alpha-2(\omega_{1}+\omega_{2}))\sigma(d_{0}+\alpha)} & = & \frac{\sigma(d_{l}-\alpha)}{\sigma(d_{l}+\alpha)}\exp\left[4\alpha\zeta\left(\omega_{1}+\omega_{2}\right)\right]\nonumber \\
 & = & \frac{\sigma^{2}(\alpha+\omega_{1}+\omega_{2})\sigma(d_{l}-\alpha)}{\sigma^{2}(-\alpha+\omega_{1}+\omega_{2})\sigma(d_{l}+\alpha)}\label{intid}\end{eqnarray}
 and this converts the quantization condition into\begin{equation}
e^{2l\zeta(\alpha)}\frac{\sigma\left(d_{0}-l_{-}+\alpha\right)\sigma\left(d_{l}-l_{+}+\alpha\right)\sigma\left(d_{0}-\alpha\right)\sigma(d_{l}-\alpha)}{\sigma\left(d_{0}-l_{-}-\alpha\right)\sigma\left(d_{l}-l_{+}-\alpha\right)\sigma\left(d_{0}+\alpha\right)\sigma(d_{l}+\alpha)}=1\,.\label{nmquant}\end{equation}
 The advantage of this form is that the $l\rightarrow\infty$ limit
(when also $l_{\pm}\rightarrow\infty$) can be carried out easily
using the asymptotic forms of the Weierstrass functions described
earlier.

As stressed earlier the solution described so far is valid for sufficiently
large $l$. As $l$ decreases from infinity $\epsilon$ increases
from zero, and the qualitative nature of the solution changes when
$\epsilon$ reaches the smaller one of the two boundary values $\Phi_{min}^{D}$.
Indeed $l_{+}$ becomes zero when $\epsilon=\Phi_{\mathrm{min}}^{D}$
and this particular $l_{0}$ is the smallest strip width when the
ground state solution has an extremum; for $l<l_{0}$ the solution
becomes monotonic and there is only one domain. In this single domain
the solution can be written as \[
x=\int\limits _{\beta\phi/2}^{\varphi_{0}}\frac{dv}{\sqrt{\sin^{2}v+C}}\quad,\textrm{or}\quad x=\int\limits _{\varphi_{0}}^{\beta\phi/2}\frac{dv}{\sqrt{\sin^{2}v+C}}\]
 where $C>-\sin^{2}\varphi_{0}$ and is determined by the width of
the strip \[
l=\int\limits _{\varphi_{l}}^{\varphi_{0}}\frac{dv}{\sqrt{\sin^{2}v+C}}\quad\textrm{or}\quad l=\int\limits _{\varphi_{0}}^{\varphi_{l}}\frac{dv}{\sqrt{\sin^{2}v+C}}.\]
 When expressed in terms of the Weierstrass's $\mathfrak{p}$ function
this solution again takes the $P_{\phi}=\mathfrak{p}(d_{0}-x)$ form
but $l$ is expressed now as $l=d_{0}-d_{l}$. The stability analysis
in the single domain leads in a straightforward manner to the quantization
condition \[
e^{2l\zeta(\alpha)}\frac{\sigma(d_{0}-\alpha)\sigma(d_{l}+\alpha)}{\sigma(d_{0}+\alpha)\sigma(d_{l}-\alpha)}=1,\qquad\quad l<l_{0}.\]

\subsubsection{(B) type solution}

Next we consider the SG model on a strip with the same Dirichlet boundary
conditions as before \[
0<\varphi_{0}<\frac{\pi}{2}\qquad,0\leq\varphi_{l}\leq\pi-\varphi_{0},\]
 but look for the (B) type solution. In this case, for \textsl{sufficiently
large} $l$-s, the solution first increases from $\Phi_{0}^{D}$ to
some maximal value $\chi=\frac{2\pi}{\beta}-\tilde{\epsilon}$ then
decreases from $\chi$ to $\Phi_{l}^{D}$ and in these two domains
it is given by slightly different expressions.

\begin{center}\includegraphics[%
  height=4cm]{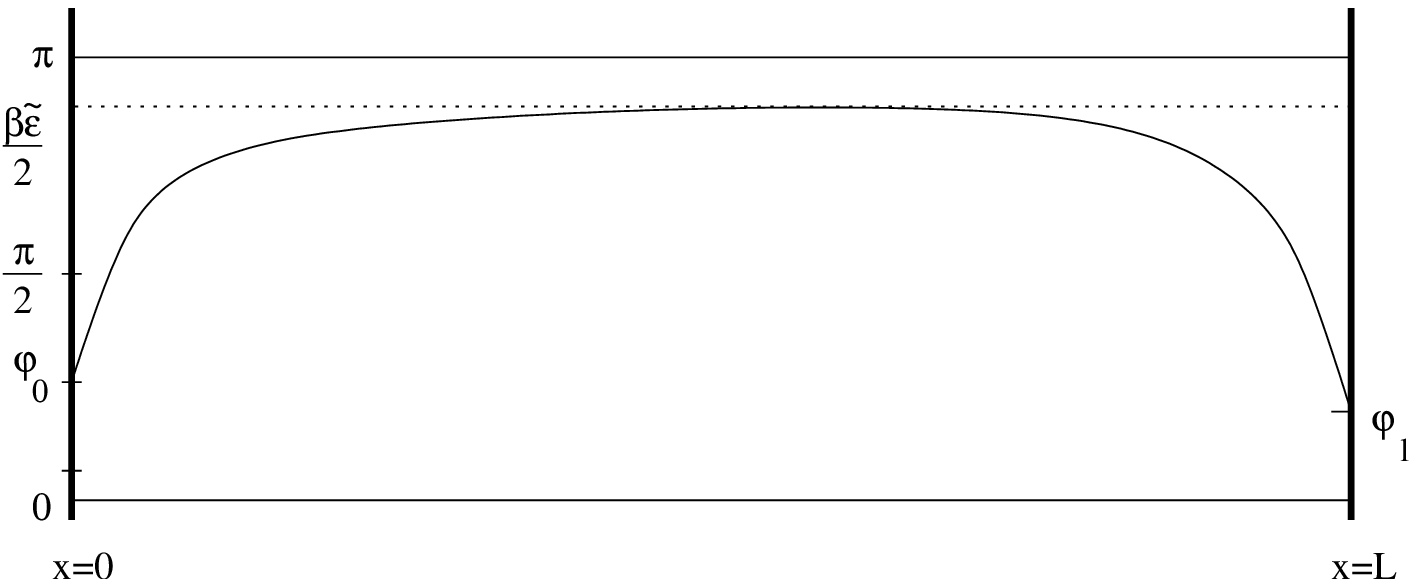}\end{center}

Since $\partial_{x}\Phi$ must vanish at the turning point the integration
constant $C$ is \[
C=-\sin^{2}(\frac{\beta\chi}{2})=-\sin^{2}(\frac{\beta\tilde{\epsilon}}{2})\leq0.\]
 Repeating the same steps as in the case of the (A) type solution
one easily obtains that this solution is expressed as \[
P_{\phi}={\mathfrak{p}}(d_{0}+x),\qquad P_{\phi}={\mathfrak{p}}(x+d_{0}+2(\omega_{1}+\omega_{2}))\]
 in the increasing (resp. decreasing) domain. The quantization condition
for the fluctuations around this background is obtained by the $d_{i}\rightarrow-d_{i}$
$i=0,\; l$ substitution in eqn. (\ref{nmquant}). This, in the classical
limit, leads to a change of sign for the $\cos\varphi_{i}$ terms.

The solution described here exists for sufficiently large $l$. As
$l$ decreases $\tilde{\epsilon}$ increases and the existence of
the solution comes to an end when $\tilde{\epsilon}$ reaches the
smaller one of the two boundary values $\Phi_{\mathrm{min}}^{D}$.
In contrast to the (A) type solution at this particular $l=l_{\mathrm{min}}$
both $l_{+}$ and $l_{-}$ remain finite and the solution does not
exist for $l<l_{\mathrm{min}}$.

To understand this consider the poles of the functions multiplying
$e^{2l\zeta(\alpha)}$ in the quantization condition as functions
of $l$. For this excited ground state there are two bound state poles,
which depend on both the boundary parameters and on $l$; the energies
of the corresponding bound states are given by (\ref{dene}) with
the $\varphi_{0}\rightarrow\varphi_{i}$ $i=0,\; l$ substitution.
It is clear that one of the $\omega^{D}$'s becomes zero at the minimal
volume, i.e. when $C$ becomes $C=-\sin^{2}\varphi_{\mathrm{min}}.$
The appearance of this zero mode clearly indicates the instability
of the underlying excited ground state solution and is in accord with
the picture described at the end of Section 4.2.

\subsection{The $-1$ topological sector: monotonic background}

Next we consider the sine-Gordon model on a strip with DD boundary
conditions characterized by the same $\varphi_{0}$ and $\varphi_{l}$
boundary parameters as before. The fundamental range of the boundary
parameters is as in (\ref{eq:fundamentalrange}), which describes
the zero charge sector. In the $-1$ topological sector, the right
boundary value of the field can obtained from the value in the zero
charge sector by the following shift: \[
\Phi(L)=\Phi_{L}^{D}-\frac{2\pi}{ \beta}\;.\]
In this case, for sufficiently large $l$ the static ground state
is the superposition of two anti-solitons rather than the superposition
of a soliton and antisoliton. 

\begin{center}\includegraphics[%
  height=4cm]{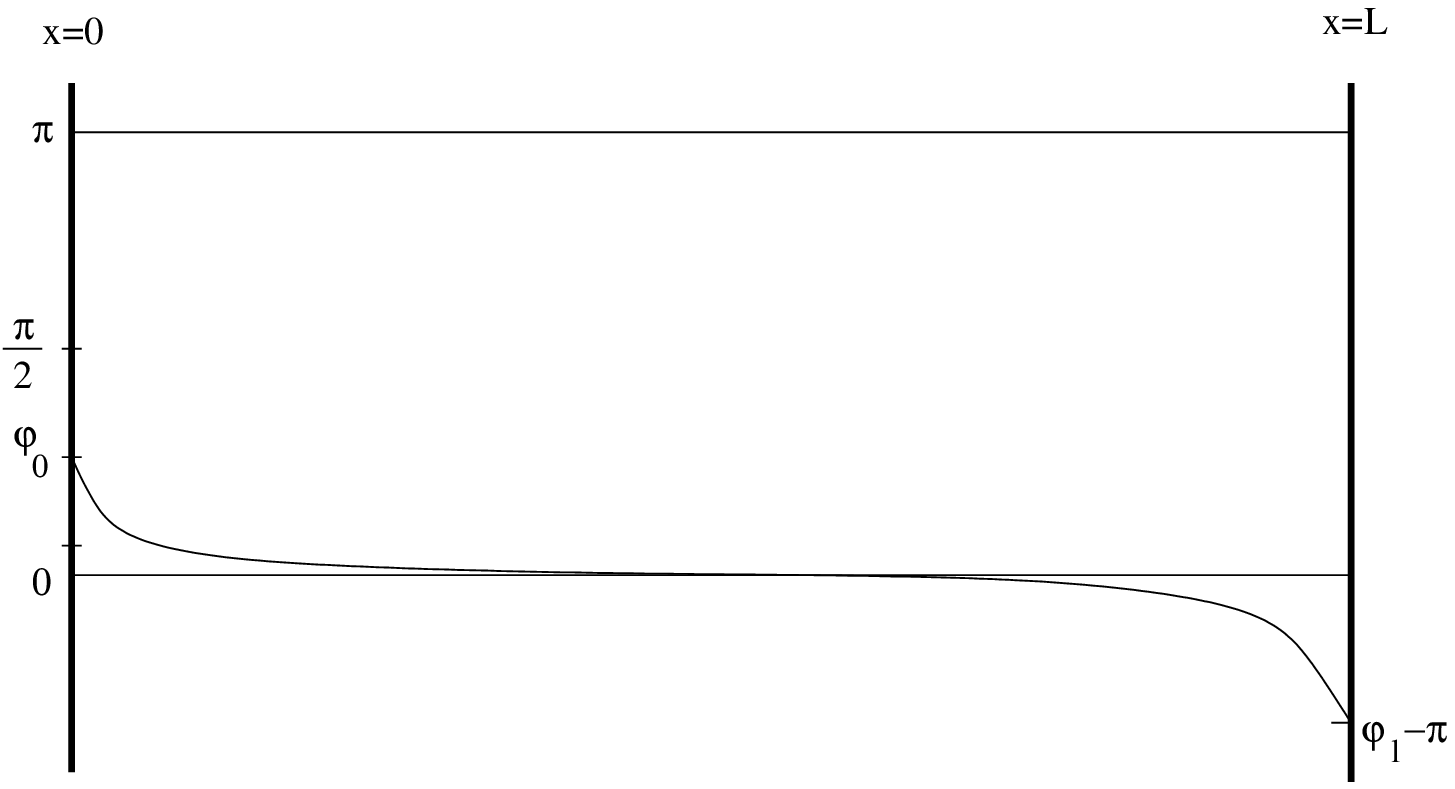}\end{center}

Because of the different signs, in the effective point particle description
the particle has to go over the maximum of $-V(\Phi)$ at $\Phi=0$.
As a consequence the integration constant $C$ has to be positive
$C>0$ and the solution becomes monotonic.

In this case there is no need to introduce $l_{\pm}$ and the two
domains but to emphasize the similarities between this case and the
previous one we define \[
l_{-}=\int\limits _{0}^{\varphi_{0}}\frac{dv}{\sqrt{\sin^{2}v+C}}=\omega_{1}+\omega_{2}-d_{0};\quad l_{+}=\int\limits _{0}^{\pi-\varphi_{l}}\frac{dv}{\sqrt{\sin^{2}v+C}}=d_{l}-(\omega_{1}+\omega_{2});\quad l=l_{-}+l_{+}.\]

Using arguments similar to the case in subsection 4.1.1, we obtain
the following result for the large volume asymptotics of the energy
(i.e. the classical limit of the scaling function):\begin{equation}
E\left(\varphi_{0},\varphi_{l},l\right)\sim\frac{4m}{\beta^{2}}\left(2-\cos\varphi_{0}+\cos\varphi_{l}+8\tan\frac{ \varphi_{0}}{2}\tan\frac{ \pi-\varphi_{l}}{2}\mathrm{e}^{-l}\right)\label{eq:energyasdirdir}\end{equation}
It is easy to see that the leading correction vanishes if $\varphi_{0}=0$
or $\varphi_{l}=\pi$, which is consistent with (\ref{eq:energyasdir0}).

In both domains the ground state solution is given by $P_{\phi}=\mathfrak{p}(x+d_{0})$.
Using this in the stability analysis leads to the quantization condition
\[
e^{2l\zeta(\alpha)}\frac{\sigma(d_{0}+\alpha)\sigma(d_{l}-\alpha)}{\sigma(d_{0}-\alpha)\sigma(d_{l}+\alpha)}=1.\]
 It is difficult to compute the $l\rightarrow\infty$ limit, as the
background does not show explicitly the presence of both anti-solitons.
To circumvent this define $\tilde{d_{0}}$ instead of $d_{0}$: \[
l_{-}=\int\limits _{0}^{\varphi_{0}}\frac{dv}{\sqrt{\sin^{2}v+C}}=\tilde{d}_{0}-(\omega_{1}+\omega_{2})=\omega_{1}+\omega_{2}-d_{0};\]
 then, using the identity (\ref{intid}), the quantization condition
can be recast as \[
e^{2l\zeta(\alpha)}\frac{\sigma(\tilde{d}_{0}-l_{-}+\alpha)\sigma(d_{l}-l_{+}+\alpha)\sigma(\tilde{d}_{0}-\alpha)\sigma(d_{l}-\alpha)}{\sigma(\tilde{d}_{0}-l_{-}-\alpha)\sigma(d_{l}-l_{+}-\alpha)\sigma(\tilde{d}_{0}+\alpha)\sigma(d_{l}+\alpha)}=1.\]
 The $C\rightarrow0_{+}$ limit, when all $l$, $l_{\pm}$ $\rightarrow\infty$
can now be computed easily.

\subsection{UV limit of ground state scaling function}

The $l\rightarrow0$ limit of ground state energy can be calculated
simply by observing that in small volume the energy functional (\ref{eq:staticenergyfunctional})
is dominated by the kinetic term \[
\int_{0}^{L}\frac{1}{2}\left( \partial_{x}\Phi\right)^{2}\]
so the ground state solution (with Dirichlet boundary conditions)
is simply\[
\Phi=\Phi_{0}^{D}+\frac{\Phi_{L}^{D}-\Phi_{0}^{D}}{L}x\]
The energy of this solution is given by \[
\frac{1}{2L}\left(\Phi_{L}^{D}-\Phi_{0}^{D}\right)^{2}\]
which agrees with the result from conformal field theory (see formula
C.7. in \cite{BPT1}). 

The fact that the ground state solution is unique in the UV limit
is consistent with our earlier observation that one of two static
solutions that exist for large volume is destabilized as the volume
decreases.

\section{Sine-Gordon model on the strip: mixed boundaries}

Now we consider a mixed three parameter family of integrable boundary
conditions on the strip: we assume Dirichlet boundary condition \[
\Phi(0)=\Phi_{0}^{D}\]
 at the lower end of the strip, while at $x=L$ the perturbed Neumann
boundary condition. Using the bulk symmetries of the theory $\Phi_{0}^{D}$
can be mapped into the fundamental range, while $\Phi_{L}$ can be
in the domain: $0\leq\Phi_{L}\leq\frac{4\pi}{\beta}$ and $M_{L}\geq0$.

In this section we consider a simpler form of the general boundary
condition (SGB): \begin{equation}
\partial_{x}\Phi|_{L}=\delta M_{L}\frac{\beta}{2}\sin(\frac{\beta\Phi_{L}}{2}),\qquad\delta=\pm1\:,\label{eq:sgb}\end{equation}
depending only on the $M_{L}$ parameter, which is assumed to be non
negative, and the positive (negative) sign in front corresponds to
choosing $\Phi_{0}^{N}=\frac{2\pi}{\beta}$ ($\Phi_{0}^{N}=0$) respectively.
Below we sometimes consider an even simpler case when $\varphi_{0}=\beta\Phi_{0}^{D}/2=0$,
called Neumann type (NT) problem.

\subsection{Classical static solutions}

At $x=L$ we may express $\hat{C}$ in terms of $\Phi_{L}\equiv\Phi(L)$
using the boundary condition: \[
\hat{C}=-\left(1-\mathcal{A}^{-2}\right)\frac{2m^{2}}{\beta^{2}}\sin^{2}\frac{\beta\Phi_{L}}{2}=\frac{2m^{2}}{\beta^{2}}C,\quad\qquad\mathcal{A}^{-1}=\frac{M_{L}\beta^{2}}{4m}.\]
 Note that $C$ is negative for $\mathcal{A}^{-1}\leq1$, i.e. as
$M_{L}$ grows slowly from zero but stays in the $[0,4m/\beta^{2})$
domain. As $(\partial_{x}\Phi)^{2}$ must be non negative, this may
impose some restrictions on the classical solution. In particular
at the $x=0$ Dirichlet end \[
\sin^{2}\varphi_{0}-\left(1-\mathcal{A}^{-2}\right)\sin^{2}\frac{\beta\Phi_{L}}{2}\geq0.\]
 Thus, for $\mathcal{A}^{-1}\leq1$ in the NT problem ($\varphi_{0}=0$)
$\Phi_{L}$ must stay either zero or $2\pi/\beta$ independently of
$L$, which implies that the solution must be the constant $\Phi(x)\equiv0$.
(Note that for $\mathcal{A}^{-1}>1$ this conclusion does not hold).

The classical solution $\phi(x)$ is obtained from (\ref{eq:genstatic})
by carefully correlating the sign in the boundary condition with that
originating in the square root: \begin{equation}
mx=\int_{\beta\Phi/2}^{\varphi_{0}}\frac{du}{R(u)}\:,\quad\delta=-1\;(\mathrm{A})\,,\qquad mx=\int_{\varphi_{0}}^{\beta\Phi/2}\frac{du}{R(u)}\:,\quad\delta=1\;(\mathrm{B})\,.\label{exint}\end{equation}
 Here $R(u)$ is the positive root \[
R(u)=\sqrt{\sin^{2}u-\left(1-\mathcal{A}^{-2}\right)\sin^{2}\frac{\beta\Phi_{L}}{2}}=\sqrt{\sin^{2}u+C},\]
 and $\phi(x)<\Phi_{0}^{D}$ in (\ref{exint} A), while $\phi(x)>\Phi_{0}^{D}$
in (\ref{exint} B)%
\footnote{In these considerations we always assume that $0<\Phi_{0}^{D}<\pi/\beta$.%
}. The equation determining $\phi(L)=\Phi_{L}$ in terms of the dimensionless
length of the strip $l=mL$ has the form \[
l=\int\limits _{\beta\Phi_{L}/2}^{\varphi_{0}}\frac{du}{R(u)},\quad\delta=-1\,\,(\textrm{A});\qquad\quad l=\int\limits _{\varphi_{0}}^{\beta\Phi_{L}/2}\frac{du}{R(u)},\quad\delta=1\,\,(\textrm{B}),\]
 in the two cases. In case (A) $l$ changes from zero to infinity
as $\Phi_{L}$ decreases from $\Phi_{0}^{D}$ to $0$. The same happens
to $l$ in case (B) as $\Phi_{L}$ increases from $\Phi_{0}^{D}$
to $2\pi/\beta$, if $\sin^{2}\varphi_{0}>1-\mathcal{A}^{-2}$. If,
however, $\sin^{2}\varphi_{0}<1-\mathcal{A}^{-2}$, then, in case
(B), $\Phi_{L}$ must be bigger than $\Phi_{\mathrm{min}}$ ($\beta\Phi_{\mathrm{min}}>\pi$),
which is obtained from \begin{equation}
\sin^{2}\varphi_{0}=(1-\mathcal{A}^{-2})\sin^{2}\frac{\beta\Phi_{\mathrm{min}}}{2},\label{stcond}\end{equation}
 and consequently the solution exists only for $l_{\mathrm{min}}<l\leq\infty$.

The energy of these classical solutions can be written in all cases
as\begin{equation}
E=\frac{4m}{\beta^{2}}\left(\left(1-\mathcal{A}^{-2}\right)\frac{l}{2}\sin^{2}\frac{\beta\Phi_{L}}{2}+ \delta\int\limits _{\varphi_{0}}^{\beta\Phi_{L}/2}R(u)du+\delta\mathcal{A}^{-1}\cos\frac{\beta\Phi_{L}}{2}\right)\,.\label{ener}\end{equation}
 Interestingly, the second term is positive, while the last term,
which is the contribution of the boundary potential, is negative for
large $l$-s for both the (A) and the (B) cases. For asymptotically
large $l$-s the energy has a simple interpretation as the sum of
the energies of the static (anti)solitons building up the ground state.
Indeed for the case (A) say it is \[
E_{\mathrm{as}}=\frac{4m}{\beta^{2}}\left(-\mathcal{A}^{-1}+1-\cos\varphi_{0}\right)\:,\]
 where the first term is the energy of a special perturbed Neumann
soliton in the left half space while the second and third terms give
the energy of a Dirichlet antisoliton in the right half space. The
asymptotic expression of the energy in the (B) case differs from this
only in the sign of the cosine term in accord with the fact that now
there is a Dirichlet soliton in the vicinity of the boundary at $x=0$.

The energy of these classical solutions as functions of the dimensionless
width of the strip $l$ was investigated numerically for a wide choice
of parameters $\Phi_{0}^{D}$ and $\mathcal{A}$. These investigations
showed that these $E(l)$-s are smooth functions that tend rather
rapidly to their asymptotic values given above; for $\delta=-1$ (case
A) this happened already at $l\sim2$ while for $\delta=1$ (case
B) at $l\sim8-10$. Furthermore these investigations revealed that
if for a given $l$ both solutions exist then always (i.e. not only
asymptotically) the energy of the (A) type solution is the smaller.
This fact is in accord with the idea that the (A) type solution is
the classical ground state and the (B) type one is some sort of classical
excitation.

\begin{figure}
\begin{center}\includegraphics{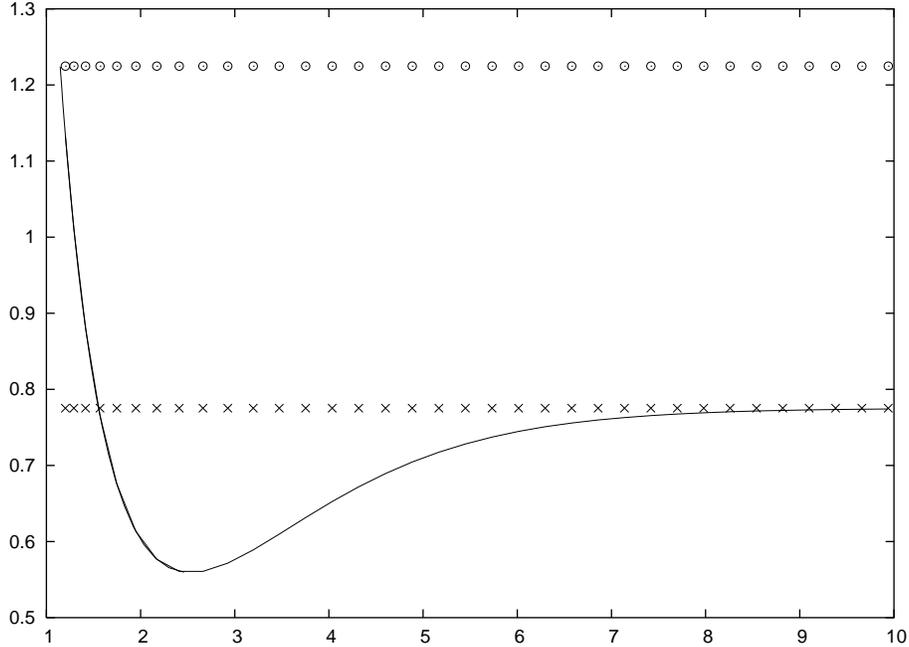}\end{center}

\caption{\label{fig:ene} $E(l)$ in the NT problem with $\mathcal{A}^{-1}=\sqrt{3/2}$ }
\end{figure}

Particularly interesting is to compare the energy $E(l)$ of the (B)
type solution of the NT problem ($\Phi_{0}^{D}=0$), when ${\mathcal{A}}^{-1}>1$
and that of the constant $\Phi\equiv0$ solution. Actually even in
this case the (B) type solution exists only for $l\geq l_{\mathrm{min}}$
where\begin{equation}
l_{\mathrm{min}}=\lim_{\beta\Phi_{L}\rightarrow0}\int_{0}^{\beta\Phi_{L}/2}\frac{du}{R(u)}=\ln\sqrt{\frac{\mathcal{A}^{-1}+1}{\mathcal{A}^{-1}-1}}\label{lmin}\end{equation}
 At $l=l_{\mathrm{min}}$ the first two terms vanish in the energy
expression (\ref{ener}) thus $E\left(l_{\mathrm{min}}\right)=\frac{4m}{\beta^{2}}\mathcal{A}^{-1}$
which is precisely the energy of the constant solution. The numerical
study showed that for $l>l_{\mathrm{min}}$ $E(l)$ is always smaller
than this particular value (see Fig. \ref{fig:ene}, where the continuous
line denotes $E(l)$, $x$ denotes its asymptotic value and the $\circ$
stands for the energy of the constant solution).

We can rewrite the solutions in terms of Weierstrass's $\mathfrak{p}(z)$
function as (\ref{Wei1}) with $d$ given by (\ref{wei2}). For $0<\mathcal{A}^{-1}<1$
the parameter $C<0$ so the roots of the cubic polynomial defining
${\mathfrak{p}}(z)$ are given by (\ref{ckn}) and the half periods
by (\ref{cknp}). In the case when $\mathcal{A}^{-1}>1$ ($C>0$)
the corresponding expressions are given by (\ref{ckp}) and (\ref{ckpp}).
Using the relation between ${\mathfrak{p}}(z)$ and Jacobi's elliptic
functions \cite{riz} the classical solution $\phi(x)$ can be given
explicitly. When $C<0$ and $\frac{1-C}{3}=e_{1}$ one has (\ref{elso}),
while for $C>0$ and $\frac{1-C}{3}=e_{2}$ we have (\ref{masodik}).
Finally we mention that in case of the (B) type solutions one has
to change $d-x$ to $d+x$ in eqns. (\ref{Wei1} ,\ref{elso} , \ref{masodik})
to get the representation in terms of elliptic functions.

\subsection{Stability analysis of the NT problem}

Before considering the more general cases we briefly describe the
stability analysis of the $\phi(x)\equiv0$ static solution. This
solution exists for all values of $\mathcal{A}^{-1}$, when $\mathcal{A}^{-1}<1$
this is the only solution, when $\mathcal{A}^{-1}>1$ and $\delta=1$
then there is also an (\ref{exint} B) type solution.

In the stability analysis we write $\Phi(x,t)=0+\xi(x)e^{-i\omega t}$
and linearize both the equation of motion \[
\left(-\frac{d^{2}}{dx^{2}}+m^{2}\right)\xi(x)=\omega^{2}\xi(x)\]
 and the boundary conditions \[
\xi(0)=0,\qquad\quad\xi^{\prime}(x)|_{L}=\delta\frac{M_{L}\beta^{2}}{4}\xi(L)\]
 around $\phi=0$. In case of the discrete spectrum we write $\omega^{2}=m^{2}(1-\epsilon^{2})$,
then the general solution of the linearized equation of motion is
\[
\xi(x)=Ae^{\epsilon mx}+Be^{-\epsilon mx}.\]
 From the two boundary conditions it is found that $A$, $B$ can
be different from zero only if \begin{equation}
\epsilon-\delta\mathcal{A}^{-1}\tanh\epsilon l=0\:,\qquad l=mL\,.\label{ntbbs}\end{equation}
 For $\delta=-1$ this equation has only the $\epsilon=0$ solution
and this implies vanishing $\xi$, thus no instability of the classical
solution. For $\delta=1$ this equation has a nontrivial solution
if $l>l_{\mathrm{crit}}=\mathcal{A}$; for $l\rightarrow\infty$ this
solution tends to $\mathcal{A}^{-1}$: $\epsilon\rightarrow\mathcal{A}^{-1}$.
If $\mathcal{A}^{-1}<1$ this implies the existence of a bound state,
while for $\mathcal{A}^{-1}>1$, when $\omega^{2}$ becomes negative
at large $l$, it means an instability. In the latter case the instability
starts at $l_{0}$ where $\epsilon(l_{0})=1$, from this condition
it is easy to show that $l_{0}=l_{\mathrm{min}}$ (cf. eqn. (\ref{lmin})).
This observation clarifies the meaning of the instability at $l>l_{0}$:
it signals the transition from the constant solution to the (B) type
one which has a lower energy.

In the semiclassical quantization \cite{Raj} the energy of the boundary
bound state associated to the nontrivial solution of eqn. (\ref{ntbbs})
is given by $\omega=m\sqrt{1-\epsilon^{2}(l)}$. Particularly interesting
is the $l$ dependence of this bound state energy, for large $l$
one finds from (\ref{ntbbs}) ($0<\mathcal{A}^{-1}<1$) \[
\omega(l)=m(\sqrt{1-\mathcal{A}^{-2}}+\frac{2\mathcal{A}^{-2}}{\sqrt{1-\mathcal{A}^{-2}}}\exp(-2\mathcal{A}^{-1}l)),\quad{\textrm{for}}\quad l\rightarrow\infty.\]
 This shows that the finite width of the strip gives an exponential
correction to the bound state energy in the infinite system. Based
on our experience with semiclassical quantization of boundary sine-Gordon
theory in infinite half space we expect \cite{KP} that the (semi)classical
limit of an exact, fully quantum computation of the finite size correction
of the boundary bound state energy would coincide with the expression
above.

In case of the continuous spectrum we write $\omega^{2}=m^{2}(1+q^{2})$
which converts the solution of the equation of motion into the sum
of plane waves. The two boundary conditions now impose the following
\lq quantization condition' on $q$: \[
1+e^{2iql}\frac{q+i\delta\mathcal{A}^{-1}}{q-i\delta\mathcal{A}^{-1}}=0.\]
 There are several things that should be remarked about this equation.
First of all the expression multiplying the exponential is nothing
but the product of the classical limits of the first breather's ($B^{1}$)
reflection factors on a ground state Dirichlet boundary with $\Phi_{0}^{D}=0$
(which is trivial $\equiv1$) and on an SGB boundary. In fact $\delta=-1$
corresponds to $B^{1}$ reflecting on the ground state boundary $\left|\right\rangle $,
while $\delta=1$ and ${\mathcal{A}}^{-1}<1$ to $B^{1}$ reflecting
on the first excited boundary $|0\rangle$ (see appendix A).

The second remark about the quantization condition concerns its relation
to the finite size correction to energy of the boundary bound state.
For this we recall that in infinite half space the bound state shows
up as a pole in the reflection amplitude at a purely imaginary value
of the rapidity and use the quantization condition to determine the
shift of this pole for large but finite $l$-s. To this end we put
$q=\sinh\theta$ and continue $\theta$ to pure imaginary values $\theta=iv$.
Furthermore, for $\mathcal{A}^{-1}<1$, introducing $\sin u=\mathcal{A}^{-1}$
the quantization condition with $\delta=1$ becomes \[
1+\exp(-2l\sin v)\frac{\sin v+\sin u}{\sin v-\sin u}=0.\]
 For large $l$ from this equation we find \[
v=u-\exp(-2l\sin u)\frac{2\sin u}{\cos u}+{\mathcal{O}}(\exp(-4l)),\]
 and this implies that the leading correction to the bound state energy
is \[
m\cos v=m\cos u+m\frac{2\sin^{2}u}{\cos u}\exp(-2l\sin u).\]
 This of course coincides with the previous result, but now it is
possible to interpret the coefficient of $l$ in the exponent and
the coefficient of the exponential correction in terms of the classical
limit of the reflection factor. In fact in these simple considerations
the classical nature was never exploited, thus had we started with
a (quantum) Bethe-Yang equation for a single particle with a (quantum)
reflection amplitude having a bound state pole at $v=u$ with residue
$iG$ we would have obtained for the bound state energy at finite
$l$ \begin{equation}
m\cos v=m\cos u-m\sin u\,\, G\exp(-2l\sin u).\label{eq:boundLuscher}\end{equation}
 We think that this expression is completely general and would give
the leading finite size correction to the energy of any boundary bound
state in the half line theory. We also expect that a derivation of
this formula must exist based on Lüscher's ideas \cite{Lusch}.

\subsection{Stability analysis}

In the general case we write $\Phi(x,t)=\phi(x)+\xi(x)e^{-i\omega t}$,
where $\phi(x)$ is the solution in (\ref{exint} A,B). Introducing
the dimensionless variable $mx\mapsto x$ the equation of motion take
the form as in (\ref{stabeq}) but with the boundary conditions: \[
\xi(0)=0,\quad\qquad\xi^{\prime}(x)|_{l}=\delta{\mathcal{A}}^{-1}\cos\left(\frac{\beta\Phi_{L}}{2}\right)\xi(l),\]
 For the (A) type solutions, using the representation in (\ref{Wei1}),
this can be rewritten as (\ref{lame}) for which the general solution
is given by (\ref{gensol}). The two boundary conditions on $\nu$
impose the following quantization condition on $p$:\begin{equation}
-\frac{\sigma(d-l-\alpha)\sigma(d+\alpha)}{\sigma(d-l+\alpha)\sigma(d-\alpha)}\mathrm{e}^{-2l\zeta(\alpha)}\frac{-\zeta(d-l)+\zeta(d-l-\alpha)+\zeta(\alpha)+Q}{-\zeta(d-l)+\zeta(d-l+\alpha)-\zeta(\alpha)+Q}+1=0\label{fulqu}\end{equation}
 where \[
Q=\delta\mathcal{A}^{-1}\cos\left.\left(\frac{\beta\Phi_{L}}{2}\right)\right|_{\delta=-1}\:.\]
 The quantization condition for the fluctuations around the (B) type
solutions is obtained by making the $d\rightarrow-d$ and $\delta=1$
substitutions in eqn. (\ref{fulqu}).

To support the classical Bethe-Yang interpretation of eqn. (\ref{fulqu})
we consider the $l\rightarrow\infty$ limit, which is implemented
by setting $\sin^{2}\frac{\beta\Phi_{L}}{2}$ (and consequently also
$C$) to zero and perform the same calculation we did in the Dirichlet
case. Since for $l\rightarrow\infty$ $Q\rightarrow-{\mathcal{A}}^{-1}$,
in the asymptotically large $l$ domain one finds from (\ref{fulqu})\begin{equation}
-\frac{ip-1}{ip+1}\frac{ip+\cos\varphi_{0}}{ip-\cos\varphi_{0}}\mathrm{e}^{2ilp}\frac{ip+\mathcal{A}^{-1}}{-ip+\mathcal{A}^{-1}}+1=0\label{ntasz}\end{equation}
 The asymptotic quantization condition for the fluctuations around
the (B) type solutions is obtained by changing the signs of $\cos\varphi_{0}$
and $\mathcal{A}^{-1}$ in (\ref{ntasz}): \begin{equation}
-\frac{ip-1}{ip+1}\frac{ip+\cos\varphi_{0}}{ip-\cos\varphi_{0}}\mathrm{e}^{2ilp}\frac{ip+\mathcal{A}^{-1}}{-ip+\mathcal{A}^{-1}}+1=0\label{ntasz2}\end{equation}
 The important difference between eqn. (\ref{ntasz}) and (\ref{ntasz2})
is that the expression multiplying the exponential in latter one admits
such poles that, for $l\rightarrow\infty$ may be interpreted as bound
states while no such pole is present in (\ref{ntasz}). The classical
energy of these bound states is given by \[
\omega_{1}=m\sin\varphi_{0},\qquad\omega_{2}=m\sqrt{1-\mathcal{A}^{-2}}.\]
 It is shown in appendix A that the expressions multiplying $e^{i2lp}$
in eqns. (\ref{ntasz},\ref{ntasz2}) are the (semi)classical limits
of the first breather's reflection amplitudes on Dirichlet and SGB
boundaries in infinite half space. 

For the (B) type solution there are potential bound state poles in
the quantization condition when $0<{\mathcal{A}}^{-1}<1$ coming from
both the Dirichlet and the SGB ends. The energy of the bound state
associated to the Dirichlet boundary is given by eqn. (\ref{dene}),
while the energy of the bound state belonging to the SGB boundary
(\ref{eq:sgb}) is given by \[
\omega^{N}=m|\cos\frac{\beta\Phi_{L}}{2}|\sqrt{1-{\mathcal{A}}^{-2}}.\]
 In this case again $\omega^{D}=0$ is obtained at the critical value
of $l$ ($l_{\mathrm{min}}$), when the static solution in question
ceases to exist since then $\Phi_{L}$ becomes $\Phi_{\mathrm{min}}$
thus $C+\sin^{2}\varphi_{0}=-(1-{\mathcal{A}}^{-2})\sin^{2}\frac{\beta\Phi_{\mathrm{min}}}{2}+\sin^{2}\varphi_{0}=0$
in view of eqn. (\ref{stcond}). The interpretation of this zero mode
is the same as in the DD case (cf. end of Section 4.2).

\section{General boundary conditions}

Now we discuss the ground state and fluctuations in boundary sine-Gordon
model when on one of the boundaries the most general, two parametric
boundary condition, while on the other the simplest Dirichlet b.c.
is imposed: \[
\partial_{x}\Phi(x)|_{0}=-M_{0}\frac{\beta}{2}\sin\left(\frac{\beta}{2}\left(\Phi(0)-\Phi_{0}^{N}\right)\right)\quad;\qquad\quad\Phi(L)=0.\]
 The integration constant appearing in \[
\frac{1}{2}\left(\partial_{x}\Phi\right)^{2}=\frac{2m^{2}}{\beta^{2}}\left(\sin^{2}\frac{\beta\Phi(x)}{2}+C\right)\]
 can be expressed in terms of the parameters characterizing the boundary
condition at $x=0$ as \[
C=\left(\mathcal{A}^{-1}\sin\left(\frac{\beta}{2}\left(\Phi(0)-\Phi_{0}^{N}\right)\right)-\sin\frac{\beta\Phi(0)}{2}\right)\left(\mathcal{A}^{-1}\sin\left(\frac{\beta}{2}\left(\Phi(0)-\Phi_{0}^{N}\right)\right)+\sin\frac{\beta\Phi(0)}{2}\right)\,.\]
 Note that $C\geq0$ is necessary to guarantee that $\Phi^{\prime}|_{L}$
is real; we assume this in what follows. Repeating the steps of the
previous derivations it is easy to see that the dimensionless width
of the strip determines now the $\chi$ parameter as %
\footnote{This form is valid if $\chi>0$; for $\chi<0$ one should interchange
the upper and lower limits of the integral.%
} \[
l=\int\limits _{0}^{\beta\Phi(0)/2}\frac{du}{\sqrt{\sin^{2}u+C}}.\]
 The advantage of this form of $C$ is that it makes obvious the asymptotic
( $l\rightarrow\infty$) limit of $\Phi(0)$: for this one must have
$C\rightarrow0$ thus \[
\Phi(0)\rightarrow\Phi(0)_{\mathrm{as}}^{\pm},\quad\tan\frac{\beta\Phi(0)_{\mathrm{as}}^{\pm}}{2}=\frac{\mathcal{A}^{-1}\sin\frac{\beta\Phi_{0}^{N}}{2}}{\mathcal{A}^{-1}\cos\frac{\beta\Phi_{0}^{N}}{2}\pm1}.\]
 In the following it is assumed that $\Phi_{0}^{N}>0$ and $\Phi(0)>0$
thus $\Phi(0)_{as}^{+}$ is used as the asymptotic limit.

The classical ground state $\phi(x)$ can be written now as \[
\frac{1-C}{3}-\sin^{2}\frac{\beta\phi(x)}{2}=\mathfrak{p}(n-x);\qquad\textrm{where}\qquad\mathfrak{p}(n)=\frac{1-C}{3}-\sin^{2}\frac{\beta\Phi(0)}{2},\]
 and the stability analysis leads in a straightforward way to the
quantization condition \[
\frac{\sigma(n-l+\alpha)\sigma(n-\alpha)}{\sigma(n-l-\alpha)\sigma(n+\alpha)}e^{-2l\zeta(\alpha)}\frac{\tilde{Q}-\zeta(n)+\zeta(n-\alpha)+\zeta(\alpha)}{\tilde{Q}-\zeta(n)+\zeta(n+\alpha)-\zeta(\alpha)}=1,\]
 where $\tilde{Q}=\mathcal{A}^{-1}\cos\frac{\beta}{2}\left(\Phi(0)-\Phi_{0}^{N}\right)$.
The large $l$ limit of this quantization condition is obtained by
replacing the Weierstrass functions with their $C\rightarrow0$ limits;
this way, after a long but straightforward algebra that involves some
surprising cancellations, one finds \[
e^{2ilp}\frac{ip-1}{ip+1}\frac{\mathcal{A}^{-1}\cos\frac{\beta\Phi_{0}^{N}}{2}-p^{2}+ip\sqrt{1+\mathcal{A}^{-2}+2\mathcal{A}^{-1}\cos\frac{\beta\Phi_{0}^{N}}{2}}}{\mathcal{A}^{-1}\cos\frac{\beta\Phi_{0}^{N}}{2}-p^{2}-ip\sqrt{1+\mathcal{A}^{-2}+2\mathcal{A}^{-1}\cos\frac{\beta\Phi_{0}^{N}}{2}}}=1.\]
Comparing with \cite{KP} one can check that this is consistent with
the classical reflection factor on a general integrable boundary.

\section{Conclusions}

The aim of this paper is to provide some intuitive picture of boundary
sine-Gordon theory on a strip by investigating the classical dynamics
of the theory. First of all, it is observed that a matching rule exists
in finite volume, meaning that not all combinations of boundary states
are allowed on a finite strip. We have seen that this can be extended
to all boundary states (not just the ground states described by static
solutions) and the resulting rule is perfectly consistent with that
observed in TCS approach.

It was also established that in general for large values of the volume
there are two different classical ground states, but under certain
conditions one of them is destabilized as the volume is decreased.
We discussed the nature of this instability and it turned out that
it corresponds to the appearance of a breather (of zero momentum quantum
number), formed by a soliton/antisoliton emitted by the left/right
boundary (or vice versa). We have shown that this behaviour perfectly
agrees with the TCS spectra, and also with the quantum Bethe-Yang
equations. In particular, for three cases we derived the asymptotic
form of the finite size correction for the energy of the ground state
(\ref{eq:energyasdir0},\ref{eq:energyasdirdir}), and we also gave
the leading correction for a boundary excited state in (\ref{eq:boundLuscher}).
It is an interesting open question to derive these (and similar formulae)
from an extension of Lüscher's argument (originally formulated for
periodic boundary conditions in \cite{Lusch}) to the strip. Work
is in progress in this direction.

The results of the present work lay the foundations for a full semiclassical
quantization of ground states on the strip, which we leave for further
investigations. Such results could be used to test and complete NLIE
descriptions proposed for the description of finite size effects,
both in the case of Dirichlet \cite{SS,ABR} and more general boundary
conditions \cite{AN}.

\subsection*{Acknowledgments.}

The authors would like to thank Ed Corrigan for illuminating discussions.
ZB is very grateful for the hospitality of APCTP during their Focus
Program 2003 where the basic ideas underlying this work were conceived.
This work was partially supported by the EC network {}``EUCLID'',
contract number HPRN-CT-2002-00325, and Hungarian research funds FKFP
0043/2001, OTKA D42209, T037674, T034299 and T043582. GT was also
supported by a Széchenyi István Fellowship, and ZB by a Bolyai János
Research Fellowship.

\subsubsection*{Note added.}

After the submission of this work a paper appeared \cite{MRS2} containing
results that partially overlap with ours.

\appendix

\makeatletter \renewcommand{\theequation}{\hbox{\normalsize\Alph{section}.\arabic{equation}}} \@addtoreset{equation}{section} \renewcommand{\thefigure}{\hbox{\normalsize\Alph{section}.\arabic{figure}}} \@addtoreset{figure}{section} \renewcommand{\thetable}{\hbox{\normalsize\Alph{section}.\arabic{table}}} \@addtoreset{table}{section} 

\makeatother

\section{The classical limit of $B^{1}$ reflection amplitudes}

In the general case the reflection factor of the first breather, $B^{1}$,
on the ground state boundary $|\rangle$ is given by \cite{gosh}\begin{equation}
R_{|\rangle}^{(1)}(\theta)=\frac{\left(\frac{1}{2}\right)\left(\frac{1}{2\lambda}+1\right)}{\left(\frac{1}{2\lambda}+\frac{3}{2}\right)}\frac{\left(\frac{\eta}{\pi\lambda}-\frac{1}{2}\right)\left(\frac{i\vartheta}{\pi\lambda}-\frac{1}{2}\right)}{\left(\frac{\eta}{\pi\lambda}+\frac{1}{2}\right)\left(\frac{i\vartheta}{\pi\lambda}+\frac{1}{2}\right)}\quad,\quad(x)=\frac{\sinh\left(\frac{\theta}{2}+i\frac{\pi x}{2}\right)}{\sinh\left(\frac{\theta}{2}-i\frac{\pi x}{2}\right)}\,.\label{b1refl}\end{equation}
 Here $\theta$ is the rapidity of $B^{1}$, $\lambda=\frac{8\pi}{\beta^{2}}-1$,
while $\eta$, $\vartheta$ are the two parameters characterizing
the bootstrap solution \cite{GZ}. ($B^{1}$'s reflection factor on
the first excited state $|0\rangle$, $R_{|0\rangle}^{(1)}(\theta)$,
is obtained from this expression by the substitution $\eta\rightarrow\bar{\eta}=\pi(\lambda+1)-\eta$
\cite{genpap}). In \cite{KP} it is shown that in the (semi)classical
limit $\lambda\rightarrow\infty$ one should scale the $\eta$ and
$\vartheta$ parameters \[
\eta=\eta_{\mathrm{cl}}(1+\lambda),\qquad\vartheta=\vartheta_{\mathrm{cl}}(1+\lambda),\]
 and keep $\eta_{\mathrm{cl}}$ and $\vartheta_{\mathrm{cl}}$ finite.
Furthermore it is also shown, that as long as $0<\eta_{\mathrm{cl}}<\frac{\pi}{2}$
the only potential pole in the ground state reflection amplitude is
out of the physical strip thus $B^{1}$ cannot create a bound state
on $|\rangle$. On the other hand $B^{1}$ can create a bound state
on $|0\rangle$ since in the same semiclassical limit $R_{|0\rangle}^{(1)}(\theta)$
has a pole at $\sim\frac{\pi}{2}-\eta_{\mathrm{cl}}$, which is in
the physical strip.

This analysis applies to both the Dirichlet case -- when the reflection
factor is obtained in the $\vartheta\rightarrow\infty$ limit of (\ref{b1refl})
-- and the simpler general (SGB) one (\ref{eq:sgb}) -- when the reflection
factor is obtained either by the $\vartheta=0$ ($\mathcal{A}^{-1}<1$)
or by the $\eta=0$ ($\mathcal{A}^{-1}>1$) substitution from (\ref{b1refl}).
Using the relation between the bootstrap and Lagrangian parameters
\cite{GZ,genpap,Zupb}, one finds in the Dirichlet case \[
\eta_{\mathrm{cl}}=\varphi_{0}=\frac{\beta\Phi_{0}^{D}}{2},\]
 thus as long as $0<\Phi_{0}^{D}<\frac{\pi}{\beta}$ there is no bound
state in the ground state reflection while there is one in the reflection
on $|0\rangle$. From the general expressions derived in \cite{KP}
it follows that in the SGB case \[
\cos\eta_{\mathrm{cl}}=\mathcal{A}^{-1}\quad\textrm{if}\quad\mathcal{A}^{-1}<1,\qquad\textrm{and}\qquad\cosh\vartheta_{\mathrm{cl}}=\mathcal{A}^{-1}\quad\textrm{if}\quad\mathcal{A}^{-1}>1.\]
 Finally starting from (\ref{b1refl}) and using $p=\sinh\theta$
instead of the rapidity one finds in the (semi)classical limit\begin{equation}
\,^{D}R_{|\rangle}^{(1)}(p)=\frac{ip-1}{ip+1}\frac{ip+\cos\eta_{\mathrm{cl}}}{-ip+\cos\eta_{\mathrm{cl}}}\label{Dlim}\end{equation}
 in the Dirichlet case while\begin{equation}
\,^{N}R_{|\rangle}^{(1)}(p)=\frac{ip+\cos\eta_{\mathrm{cl}}}{-ip+\cos\eta_{\mathrm{cl}}}\qquad\mathrm{or}\qquad\,^{N}R_{|\rangle}^{(1)}(p)=\frac{ip+\cosh\vartheta_{\mathrm{cl}}}{-ip+\cosh\vartheta_{\mathrm{cl}}}\label{Nlim}\end{equation}
 in case of the SGB boundary. The expressions for the limiting values
of the reflection amplitudes on the first excited state are obtained
by making the substitution $\eta_{\mathrm{cl}}\rightarrow\pi-\eta_{\mathrm{cl}}$,
which amounts to changing the signs of the cosine terms in eqns. (\ref{Dlim},\ref{Nlim}).
This completes the interpretation of the expressions appearing in
the asymptotic quantization conditions (\ref{ntasz},\ref{ntasz2}).

\section{Numerical analysis of the quantization condition}

In this appendix we investigate numerically the difference between
the predictions of the exact finite volume quantization condition
and the asymptotic ($l\rightarrow\infty$) one to get an estimate
on the accuracy of the latter. In the analysis we consider the DD
boundary conditions with $\varphi_{l}=0$ and the (+) case only.

In the procedure (with a fixed $\varphi_{0}$) the input parameter
is $C$, from this $l$ and $d$ are determined first, according to
(\ref{lasfC}) and (\ref{wei2}), then, using them in the logarithmic
version of (\ref{fulqu})\begin{equation}
2l\zeta\left(\alpha_{n}\right)+\log\left(\frac{\sigma(d-\alpha_{n})\sigma(d-l+\alpha_{n})}{\sigma(d+\alpha_{n})\sigma(d-l-\alpha_{n})}\right)=2i\pi n\,,\label{exl}\end{equation}
 $\alpha_{n}$ is obtained, and finally the square of quantized momentum
is found from (\ref{qmom}). The asymptotic quantization condition
is obtained from (\ref{exl}) by replacing the Weierstrass functions
with their $C\rightarrow0$ ($l\rightarrow\infty$) limits described
earlier; this way one obtains\begin{equation}
2ilp_{n}+\log\left(\frac{(ip_{n}-1)\left(ip_{n}+\cos\varphi_{0}\right)}{(ip_{n}+1)\left(ip_{n}-\cos\varphi_{0}\right)}\right)=2i\pi n\,.\label{asl}\end{equation}
 We carried out the numerical analysis with $\varphi_{0}=\pi/6$;
the results for $n=1$ are summarized in the table below:

\begin{center}\begin{tabular}{|c|c|c|c|c|}
\hline 
$C$&
 $l$&
 $p^{2}$&
 $p_{\textrm{as}}^{2}$&
 $(\pi/l)^{2}$\tabularnewline
\hline
0.4&
 0.757175&
 17.079765&
 16.876796&
 17.214972\tabularnewline
\hline
0.1&
 1.287219&
 5.843204&
 5.774261&
 5.956549\tabularnewline
\hline
0.01&
 2.376994&
 1.680518&
 1.671835&
 1.746801\tabularnewline
\hline
0.001&
 3.523447&
 0.760057&
 0.759118&
 0.794994\tabularnewline
\hline
0.00081&
 3.628742&
 0.716545&
 0.715781&
 0.749527  \tabularnewline
\hline
\end{tabular}\end{center}

Here $p_{\mathrm{as}}^{2}$ is the momentum squared obtained from
eqn. (\ref{asl}) and the last column shows the momentum squared of
a free particle. These result indicate that the predictions of the
asymptotic quantization condition are very good estimates of the exact
finite volume ones even at intermediate strip widths. (Note that here
$l$ is measured in units of the fundamental scalar particle/first
breather mass $l=mL$ while in TCSA studies it is usually measured
in units of the soliton mass $l_{\textrm{TCSA}}=ML=(M/m)mL=(M/m)l$).

\end{document}